\documentclass[
    aps,prx,twocolumn,
    reprint,
    superscriptaddress,
    floatfix,
    amssymb,
    longbibliography
]{revtex4-1}

\usepackage{amsmath,xfrac,array}
\usepackage{amsfonts}
\usepackage{graphicx}
\usepackage{bm}
\usepackage[utf8]{inputenc}
\usepackage{nicefrac}
\usepackage[normalem]{ulem}
\usepackage{xcolor}

\definecolor{linkColor}{rgb}{0,0.3,0.7}
\usepackage{hyperref}
\hypersetup{colorlinks=true,
            allcolors=linkColor,
            pdfborder={0 0 0},
            pdfencoding = auto
            }

\definecolor{darkblue}{RGB}{0,0,200}
\definecolor{darkred}{RGB}{200,0,0}
\definecolor{myGreen}{RGB}{19,132,23}

\usepackage{bm}
\usepackage{tabularray}
\usepackage{placeins}

\definecolor{myGreen}{RGB}{19,132,23}

\usepackage{mathtools}
\newtagform{normalsize}[\normalsize]{\normalsize(}{\normalsize)}

\begin{document}
\usetagform{normalsize}
\preprint{APS/123-QED}

\title{Acoustic signaling enables collective perception and control in active matter systems}

\author{Alexander Ziepke}
\author{Ivan Maryshev}%
\affiliation{%
Arnold Sommerfeld Center for Theoretical Physics and Center for NanoSciences, Ludwig-Maximilians-Universit\"{a}t M\"{u}nchen, Theresienstra{\ss}e 37, 80333 M\"{u}nchen, Germany
}%

\author{Igor S. Aranson}
 \email{isa21@psu.edu}
\affiliation{
Department of Biomedical Engineering, Pennsylvania State University, University Park, PA 16802, Pennsylvania, USA
}%

\author{Erwin Frey}
\email{frey@lmu.de}
\affiliation{%
Arnold Sommerfeld Center for Theoretical Physics and Center for NanoSciences, Ludwig-Maximilians-Universit\"{a}t M\"{u}nchen, Theresienstra{\ss}e 37, 80333 M\"{u}nchen, Germany
}%
 \affiliation{
 Max Planck School Matter to Life, Hofgartenstra{\ss}e 8, 80539 M\"{u}nchen, Germany
}%

\date{\today}

\begin{abstract}
Emergent cooperative functionality in active matter systems plays a crucial role in various applications of active swarms, ranging from pollutant foraging and collective threat detection to tissue embolization.
In nature, animals like bats and whales use acoustic signals to communicate and enhance their evolutionary competitiveness. Here, we show that information exchange by acoustic waves between active agents creates a large variety of multifunctional structures. In our realization of collective swarms, each unit is equipped with an acoustic emitter and a detector. The swarmers respond to the resulting acoustic field by adjusting their emission frequency and migrating toward the strongest signal. We find self-organized structures with different morphology, including snake-like self-propelled entities, localized aggregates, and spinning rings.
These collective swarms exhibit emergent functionalities, such as phenotype robustness, collective decision-making, and environmental sensing. For instance, the collectives show self-regeneration after strong distortion, allowing them to penetrate through narrow constrictions. Additionally, they exhibit a population-scale perception of reflecting objects and a collective response to acoustic control inputs. Our results provide insights into fundamental organization mechanisms in information-exchanging swarms.  They may inspire design principles for technical implementations in the form of acoustically or electromagnetically communicating microrobotic swarms capable of performing complex tasks and concerting collective responses to external cues.
\end{abstract}

\maketitle


\section{\label{sec:level1}Introduction}
What are the most distinct markers of living systems? What makes them so different from the inanimate world? These are their ability to move (locomotion), consume energy (metabolism), process information, and form multicellular aggregates. The onset of collective behavior among simple interacting units is a central paradigm in nonequilibrium physics and an opportunity for materials science and microrobotics \cite{li2022self,gardi2022microrobot,ori2023observation}. 
Biological systems exhibit diverse signaling strategies and mutual synchronization, giving them an evolutionary advantage \cite{sales2012ultrasonic,couzin2018synchronization}. For instance, social amoeba use cell-to-cell signaling through the emission of cyclic adenosine monophosphate (cAMP) concentration waves and chemotaxis to aggregate under starvation  \cite{parent1999cell}, 
insects rely on sound to coordinate the formation of cohesive swarms \cite{gorbonos2016long,sinhuber2021equation}, bats and whales use ultrasound sonar for communication, navigation, and hunting \cite{barrett-lennard1996mixed,boonman2019benefits}.
To what degree can one use simple information processing capabilities to design self-organized functional aggregates from simple building blocks or to create swarms of active agents performing elaborate tasks collectively \cite{croon2022insect}?
Biological systems have mastered complex functionality and environmental adaptation through evolution and 
 self-organization: the tendency of simple units  (e.g., molecules, colloids, cells) to form hierarchical functional superstructures \cite{whitesides2002self,wang2015from}.  Out-of-equilibrium self-organization opens the way to sophisticated aggregated states with many levels of functionality akin to living systems  \cite{whitesides2007dont,zampetaki2021collective}. Imagine synthetic bottom-up systems capable of communicating, making decisions, adapting, and even repairing damages in their collective structures. 
 However, currently, these features are mostly lacking in simple microrobots \cite{dong2020controlling,yang2022survey}, and enabling communication and self-organization in synthetic swarm-like systems is so far perceived as nothing more than science fiction \cite{schatzing2006swarm,crichton2008prey}.

Chemical signaling and emergent self-organization have been extensively studied in various biological and synthetic systems \cite{Taga2003Chemical}, such as social amoebas~\cite{Singer2019Oscillatory}, chemically interacting colloidal particles \cite{altemose2017chemically}, and pheromone-driven social insects \cite{Czaczkes2015Trail}. While effective in facilitating coordinated behaviors, this form of communication is inherently limited by localized diffusive spreading of information and the relaying of signals by distributed agents \cite{ziepke2022multi}. As a result, information propagation is comparably slow and remains constrained to regions where agents are present. In contrast, signaling via acoustic and electromagnetic waves, readily accessible to technological systems like microrobots, offers greater versatility and range. Yet it remains largely underexplored in the context of collective behaviors and self-organization.

To examine the potential of acoustic signaling in  technical systems, we enhance the communication capability of individual units (swarmers) through rapid signal exchange by propagating acoustic, electromagnetic, or surface waves.  
Each self-propelled swarmer is equipped with an ``on-board'' oscillator, broadcasting and detecting acoustic signals \cite{strogatz1993coupled,rosenblum2001phase}. 
Synchronizing internal oscillator states enhances the swarms' cohesiveness, multifunctionality, and robustness to heterogeneities or disorder. 
The self-organization of a collective swarm is guided by the following principles: the swarmers respond to the common acoustic field by synchronizing their broadcast frequencies and migrating towards the strongest signal. 

Our computational studies reveal the spontaneous formation of a plethora of self-organized structures with different morphology,
including snake-like self-propelled objects, localized aggregates, and closed, rotating rings.
Importantly, these structures exhibit emergent functionalities like phenotype robustness, collective decision making, and environmental sensing.
Some structures, like snakes, exhibit shape memory and self-regeneration: after a strong distortion, they are able to recover earlier phenotypes. 
  Akin to an octopus escaping from a cage through a tiny hole, the snake swarm can squeeze through a narrow constriction and reassemble behind it. These results suggest new design principles and control strategies for
multifunctional synthetic swarms which could be relevant for various applications, e.g., for pollutant foraging \cite{liu2020hydrogel,chang2022nature,urso2023smart}, threat detection \cite{hu2011school}, and tissue embolization \cite{law2022microrobotic,oral2023invivo}.

Moreover, our approach enriches the traditional scope of active matter: the onset of collective behavior emerging in the system of interacting self-propelled particles 
\cite{gompper2020motile,aranson2022bacterial}. In addition to alignment interactions in active systems \cite{chate2020dry}, our agents communicate via acoustic waves and synchronize their intrinsic states. The long-range coupling brings this system closer to the celebrated Kuramoto model of globally coupled oscillators \cite{acebron2005kuramoto}.  Notably, our agents dynamically reconfigure the coupling due to the self-organized motion.
Contrary to ``swarmalator models'' \cite{okeeffe2017oscillators,ceron2023diverse},
we disentangle the spatial orientation of agents from their internal communication state, making the system more suitable for synthetic implementations such as in microrobotic swarms.

\section{Model description}

\subsection{Agent-based model}
\label{sec:model_agents}

\begin{figure}[thb]
\centering
\includegraphics[width=\linewidth]{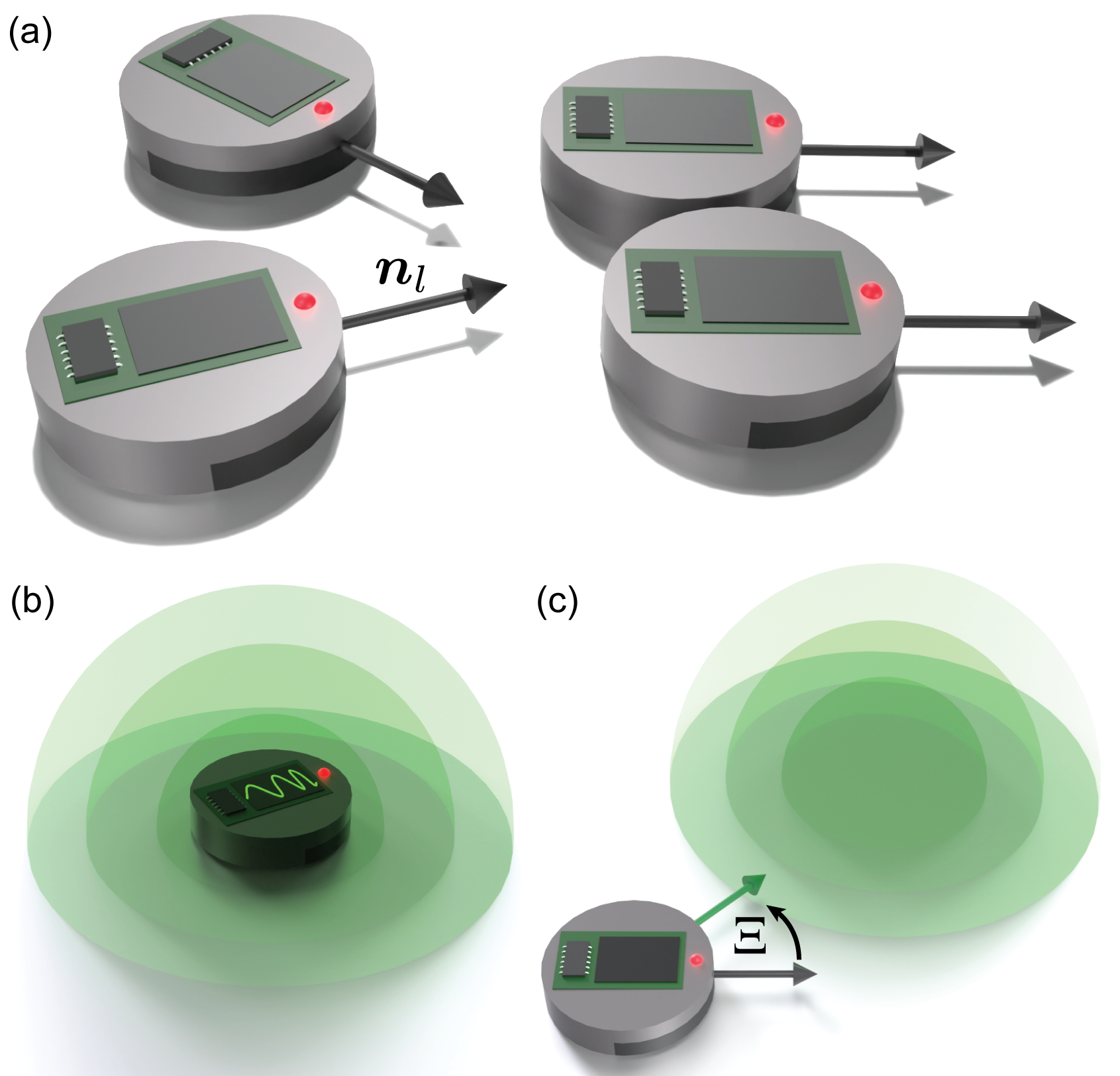}
\caption{\label{fig:schematics}{\bf Acoustically communicating active matter.} (a) Schematics of self-propelled swarmers with polar alignment. Arrows indicate the swarmers' direction of motion, $\bm{n}_l$. (b) Individual swarmers possess an internal oscillator controlling their acoustic emissions. In turn, their oscillatory states can synchronize via the acoustic field. (c) Swarmers align towards higher sound amplitudes with acoustic susceptibility $\Xi$.}
\end{figure}

In this study, we investigate the dynamics of self-propelled polar agents coupled through wave interactions, using acoustic waves as a concrete example.
For simplicity, we assume that the swarmers move in a two-dimensional plane (their ``habitat''), while sound wave propagation occurs in three spatial dimensions, leading to a realistic non-local acoustic coupling.
Specifically, we consider a system of $N$ acoustically interacting self-propelled particles (swarmers); see Fig. \ref{fig:schematics}. 
The swarmers move persistently along their intrinsic orientation ${\bm{n}_l=\left(\cos\varphi_l,\sin\varphi_l\right)^\mathrm{T}}$ at a constant speed $v_0$ with two-dimensional orientation angle $\varphi_l$. 
Within an interaction radius $r_c$, they align their direction of motion with the one of their neighbors (Fig.\@ \ref{fig:schematics}a). 
A repulsive force $f_{lj}$ ensures hard-core repulsion between agents when they come within a distance of two agent radii, $r_\mathrm{p}$. 
The dynamics of the agent's orientation angle $\varphi$ are governed by polar alignment with neighbors at a rate $\Gamma$, as well as by alignment with the amplitude of the common sound field, i.e., ${\varphi_s = \mathrm{angle}(\nabla\lvert u \rvert)}$, at a rate $\Xi$.
The adaptation of orientation $\varphi_l$, Eq.~\eqref{m_phi}, is assumed to be error-prone. 
Therefore, a zero-mean Gaussian white noise $\xi_l$ is added to the angular dynamics as a perturbation. 
In summary, the agents' position and orientation change according to
\begin{subequations}
\label{eq:agents}
\begin{align}
    \frac{\mathrm{d}\bm{r}_l}{\mathrm{d}t}&=v_0\bm{n}_l+\sum_{j[r_{lj}<2r_{\mathrm{p}}]}f_{lj}\,{,} \label{m_rv}\\
    \frac{\mathrm{d}\varphi_l}{\mathrm{d}t}&=-\Gamma\sum_{j[r_{lj}<r_c]}\frac{\sin\left(\varphi_l-\varphi_j\right)}{\lvert\bm{r}_l-\bm{r}_j\rvert}\nonumber\\
    &\quad+\Xi\sin\left(\varphi_s-\varphi_l\right)+\xi_l\,. \label{m_phi}
\end{align}
We consider agents that are equipped with an internal oscillator continuously emitting sound waves like a loudspeaker.
For simplicity, we model the oscillation using a generic form near a supercritical Hopf bifurcation, represented by a Stuart-Landau oscillator \cite{strogatz2018nonlinear,aranson2002world},
\begin{eqnarray}
    \partial_t a_l(t)&=\left(1+i\omega\right)a_l-\left(1+ib\right)\lvert a_l\rvert^2a_l+\lambda u(\bm{r}_l,t)\,.
    \label{sle}
\end{eqnarray}
Here, $\omega$ and $b$ describe the linear, respectively, non-linear coupling of frequencies and amplitudes, and $\lambda$ determines the coupling rate to the acoustic field $u(\bm{r},t)$. Thus, the intrinsic oscillator is affected by acoustic waves from other swarmers, creating acoustic feedback (Fig.\@ \ref{fig:schematics}b). Via emitting acoustic waves, the swarmers establish a dynamic signaling landscape (\textit{soundscape}) to which they adapt their intrinsic oscillation state with respect to their individual baseline frequency $\omega$. The dynamics of the common acoustic field $u(\bm{r},t)$ generated by all constituting agents is given by
\begin{eqnarray}
    \frac{1}{c^2}\partial_t^2 u(\bm{r},t)&=\nabla^2 u +\sum_{j=1}^N w\left( \bm{r}-\bm{r}_j\right)a_j\delta(z)\,.
    \label{we}
\end{eqnarray}
\end{subequations}
The soundscape evolution is described by the wave equation in three spatial dimensions with the swarmers' oscillating membranes as sources $w(\bm{r})$ of the sound field, Eq.\@ (\ref{we}). 
The function $w$ specifies the shape of the agent (i.e., round). 
The speed of sound is denoted by a parameter $c$ and is assumed to be large compared to the agent velocity $v_0$. 
The $\delta(z)$ function in Eq.\@ (\ref{we}) stipulates that all agents are confined to the ${z=0}$ plane.
As stated above, in the presence of an established acoustic field, agents align their motion toward regions of higher acoustic field amplitudes with susceptibility $\Xi$ (Fig.\@ \ref{fig:schematics}c).
To integrate numerically the dynamics of the acoustically interacting self-propelled polar agents, we solve the discrete stochastic equations, Eqs.\@ (\ref{m_rv}), (\ref{m_phi}), and the agents' oscillatory dynamics, Eq.\@ (\ref{sle}), with a forward Euler-Maruyama method at fixed time steps $dt$ \cite{kloeden1992stochastic}.  
The resulting acoustic field is calculated from Eq.\@ (\ref{we}) for the quasi-stationary case in the limit of large sound velocities $c\gg v_0$ using an inverse Fourier transform of the analytic expression
\begin{eqnarray}
\label{eq:sound_sol}
   \tilde{u}_{\bm{k}}=\frac{\tilde{g}_{\bm{k}}}{2c^2\sqrt{k_x^2+k_y^2-\omega_u^2/c^2}} \,,
\end{eqnarray}
with Fourier-transformed acoustic source contributions $\tilde{g}_{\bm{k}}$, wave vector $\bm{k}=\left(k_x,k_y\right)^T$, and dominant field frequency $\omega_u$; see also Appendix \ref{sec_app:static}.
We discretize the entire habitat domain into Fourier modes down to length scales of the order of a single agent diameter. 
Subsequently, we integrate the states of the intrinsic oscillators by incorporating the collectively established acoustic field, see Appendix \ref{app:num} for details on the numerical integration of the model equations.

\subsection{Continuous field equations}

We complement the agent-based discrete description of the system with corresponding continuous field equations.
The phenomenological equations for the agent density $\rho(\bm{r},t)$ and the particles' polar orientation field $\bm{p}(\bm{r},t)$ in the two-dimensional habitat $\bm{r}\in\mathbb{R}^2$ read 
\begin{subequations}
\label{eq:field}
\begin{align}
    \partial_t\rho(\bm{r},t)&=-v_0\partial_i p_i+\mu\nabla_\text{2D}^2\rho\,,\label{eq:field_rho} \\
    \partial_t p_i(\bm{r},t)
    &=\sigma\left(\rho-1\right)p_i-\delta p_j p_j p_i +\kappa\nabla_\text{2D}^2p_i
    \nonumber\\
    &\quad-\chi p_j\partial_j p_i-\frac{v_0}{2}P'(\rho)\partial_i\rho+\rho\,\Xi\,\partial_i\lvert u\rvert^2
    \label{eq:field_pol} \, ,
\end{align}
where the spatial derivatives $\partial_i$ with ${i\in\left\{x,y\right\}}$ and the Laplacian $\nabla_\text{2D}$ refer to the two-dimensional habitat. 
Field theories with similar contributions have been developed for a range of active matter systems \cite{Toner1995Long,Julicher2018Hydrodynamic,Shaebani2020Computational}. 
These theories are particularly relevant for studying biological and experimental polar and nematic active systems \cite{Marchetti2013Hydrodynamics,chate2020dry}, such as microtubule-kinesin mixtures \cite{Serra2023Defect}  and the actin motility assay \cite{Huber2018Emergence}. Recently, in the latter case, a combination of computational and field-theoretical approaches has provided insights into the coexistence of nematic lanes and defects in the motility assay \cite{Kruger2023Hierarchical}. In addition to phenomenological models based on symmetry considerations, hydrodynamic field equations can be derived from Smoluchowski~\cite{Baskaran2008Hydrodynamics} and Fokker-Planck equations~\cite{Romanczuk2012Mean}, or through the Boltzmann-Ginzburg-Landau framework \cite{aranson2005pattern,Bertin2006Boltzmann,Bertin2009Hydrodynamic,peshkov2012nonlinear}. 
The latter has been successfully applied, for instance, to inelastically aligning microtubule systems \cite{aranson2005pattern,Maryshev2018Kinetic,MaryshevSM2019} and to actin dynamics in motility assays \cite{Denk2016Active}. 
The approach can also be used to describe biopolymers systems driven by interactions mediated by molecular motors \cite{DeLuca2024Supramolecular}, as well as the collective behavior of chemotactic polar active agents \cite{ziepke2022multi}, among others.

The rationale behind the above set of continuum equations is as follows: The dynamics of the density is described by a diffusion-advection equation, Eq.~\eqref{eq:field_rho}, where alongside spatial diffusion with diffusion coefficient $\mu$, the agents' self-propulsion induces an advection of density with velocity $v_0$ along the direction of polar orientation. 
The dynamics of the density-weighted polar orientation $p_i$ is governed by a third-order polynomial in $p_i$, describing an isotropic to polar order transition at critical density ${\rho_c \equiv 1}$. 
For low densities, ${\rho < \rho_c}$, angular diffusion of agents dominates and the system favors the isotropic state. 
In contrast, in denser regions, ${\rho > \rho_c}$, an increased polar alignment between agents induces polar order.
Additionally, we consider elastic contributions to the polar field ($\sim\kappa$), which originate from polar alignment of neighboring agents, and self-advection ($\sim \chi$) of the polar director along the direction of the agent motion $p_i$.
Moreover, a pressure-like contribution ($\sim P'(\rho)$, with a prime denoting the derivative w.r.t.\@ the density, App.~\ref{app:num}) implements the assumed finite volume of the agents and ensures a maximum density of swarmers. 
Finally, similar to chemotactic models \cite{Romanczuk2008Beyond}, the polar orientation is coupled to gradients in the signaling field. Here, it is given by the acoustic field amplitudes $\lvert u\rvert^2$ and the acoustic susceptibility $\Xi$ controls the agents' alignment strength. 

The continuum equations for the acoustic field and the oscillatory states of the agents are obtained by coarse-graining their agent-based representations. 
Similar to the agent-based model, swarmers with density $\rho(\bm{r},t)$ and state $a(\bm{r},t)$ act as sources of the acoustic field. 
In the continuum description, the discrete contributions from individual agents become a continuous source term, weighted by the agent density field $\rho(\bm{r},t)$.
The rest of the wave equation, Eq.~(\ref{we}), for the propagation of acoustic signals with sound velocity $c$ remains unchanged. 
The oscillatory states of the agents are transformed into a continuous field, $a(\bm{r},t)$.  
Coarse-graining also introduces a diffusion-like contribution to this field ($\sim\mu$), corresponding to the positional diffusion of agents. 
Additionally, since the oscillatory states are tied to the individual agents, the state field is advected with the agent velocity ($\sim v_0$), similar to the analysis performed in Ref.~\cite{ziepke2022multi} for chemical agent states, with the limit cycle oscillations now distributed across the entire domain.
As in the discrete model, the synchronization of oscillatory states is captured through coupling to the acoustic field ($\sim\lambda$).
Altogether, the continuum representation of the equations reads
\begin{align}
\frac{1}{c^2}\partial_t^2u(\bm{r},t)&=\nabla^2u+a\,\rho\,\delta(z)\,, \label{eq:sound_cont}\\
    \partial_t a(\bm{r},t)&=\mu\nabla_{\text{2D}}^2a+\left(1+i\omega\right)a-\left(1+ib\right)\lvert a\rvert^2a\nonumber\\
    &\quad-v_0\frac{p_j}{\rho}\partial_j a+ \lambda \,u\,.
\end{align}
\end{subequations}

For homogeneous densities, ${\rho=\rho_0}$, one can integrate Eq.\@ (\ref{eq:sound_cont}) in Fourier space which introduces a long-range coupling to the state field $a$. 
This transforms the system into a non-locally (acoustically) coupled version of the complex Ginzburg-Landau equation (CGLE) \cite{aranson2002world}, as shown in App.\@ \ref{app:cgle}. 
The emergence of a non-local coupling of oscillators through the acoustic field underscores potential advantages that acoustic signaling offers to active matter systems.
In the case of homogeneous density, $\rho=\rho_0$, we observe that acoustic interactions accelerate the coarsening of phase defects compared to the CGLE, indicating the long-range interaction between the emergent spiral cores, see Appendix Fig.\@ \ref{fig:defect_coarsening}. Moreover, this coarsening process halts at a characteristic length scale, where acoustic interactions prevent further attractive forces between phase defects, highlighting the potential of long-range communication and synchronization between self-organizing collectives.
These effects are crucial for understanding the emergence of the self-organized structures and their mutual interactions, which we will explore in the following sections.

\section{Self-organized states}
\label{sec:States}

As described above, acoustic coupling between agents generates long-range interactions which may lead to intriguing forms of self-organization and novel functional structures. 
In this section, we examine the properties and distinct acoustic signatures of these emergent structures and determine the conditions under which they occur. 
We start by employing discrete agent-based simulations of Eqs.\@ (\ref{eq:agents}) and extend our analysis with large-scale numerical simulations of the continuous field equations, Eqs.\@ (\ref{eq:field}), in the latter part of this section.

\subsection{Agent-based simulations}
Acoustic signaling enables the swarmers to communicate, navigate, and assemble. 
Through acoustic communication, the agents' oscillatory states can locally synchronize, enhancing local sound amplitudes. As agents move towards these high-amplitude regions, self-organized collective states with distinct properties emerge. 

\subsubsection{Acoustic interactions enable aggregation}
We assess the specific role of acoustic coupling in the initial aggregation process and the formation of collective states. To achieve this, we consider intermediate agent densities—low enough to remain below the critical threshold for motility-induced phase separation, yet high enough to ensure sufficiently strong acoustic interactions between agents.
\begin{figure*}
    \centering
    \includegraphics[width=\linewidth]{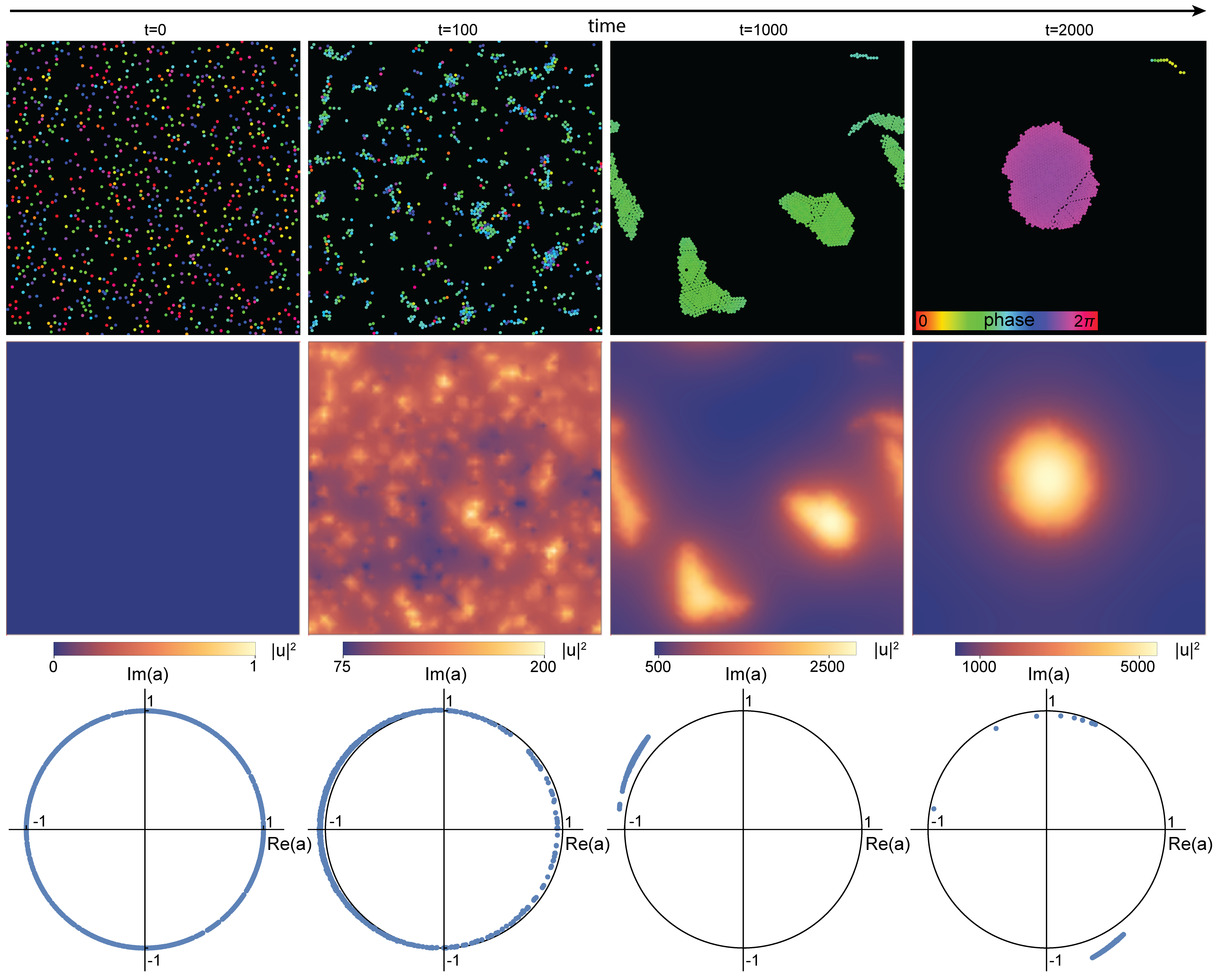}
    \caption{\textbf{Acoustically mediated formation of aggregates.} Temporal evolution of active agents (top), emitted acoustic field amplitudes (middle), and distribution of the oscillatory states (bottom). The agents synchronize their phases (color code) over time as mediated by the acoustic interactions. This, in turn, leads to higher acoustic field amplitudes $\lvert u\rvert$, further amplifying the synchronization of the agents' oscillatory states. The level of synchronization in the system can be extracted from the position of the oscillatory states $a$ in the complex plain (bottom row). Initially, at $t=0$, the agents' phases are uniformly distributed.  At later times clusters of partially synchronized agents emerge that also exhibit higher oscillation amplitudes as they deviate from the unit circle. The simulation domain has a side length of 50 units and parameters are given in Tab.~\ref{tab:partPars}.}
    \label{fig:aggregation}
\end{figure*}
Starting from an initial, spatially homogeneous distribution of agents with uniformly distributed phases (Fig.~\ref{fig:aggregation}, $t=0$), the agents emit acoustic signals with uncorrelated phases. During this initial stage, the agents lack synchronization and acoustic field amplitudes are comparably low. However, upon establishing a mutual acoustic field, the agents begin synchronizing their phases through the acoustic coupling. This synchronization is represented by an increasing heterogeneity in the distributions of oscillatory states $a$ (Fig.~\ref{fig:aggregation}, $t=100$). In turn, the formation of locally synchronized collective oscillations gives rise to regions with higher acoustic field amplitudes that emerge as self-organized aggregation centers. Agents turn towards these regions and successively synchronize their oscillatory phases to the locally dominating signals. After some time, more and more clusters form and merge with neighboring aggregates through mutual synchronization until a single persistent collective aggregate remains (Fig.~\ref{fig:aggregation}, $t=1000$ and $t=2000$).
Altogether, this shows that the acoustic interactions among active agents play a crucial role in the aggregation process. Initially, these interactions enable the synchronization of the agents' oscillatory states. Once synchronized, the agents emit acoustic waves in unison with larger collective sound amplitudes that guide the active motion of agents toward these regions and thereby enable aggregation. 
The emergent attraction between agents, facilitated by acoustic interactions, differs fundamentally from conventional chemical interactions, which rely on the diffusive transport of chemical concentrations or chemical wave propagation \cite{Romanczuk2008Beyond,Liebchen2018Synthetic,ziepke2022multi}. In the model, described by Eqs.\@ (\ref{eq:agents}), agent attraction arises through the synchronization of oscillations enabled by physical acoustic interactions. This synchronization is inherently bi-directional, as it occurs through the superposition of acoustic signals, allowing agents to dynamically influence each other on short time scales and over long distances. 
In the following, we will explore which different collective states form in the system and how the aggregates' behavior depends on key parameters of the model.

\subsubsection{Self-organized collective states}
\begin{figure*}
\centering
\includegraphics[width=\linewidth]{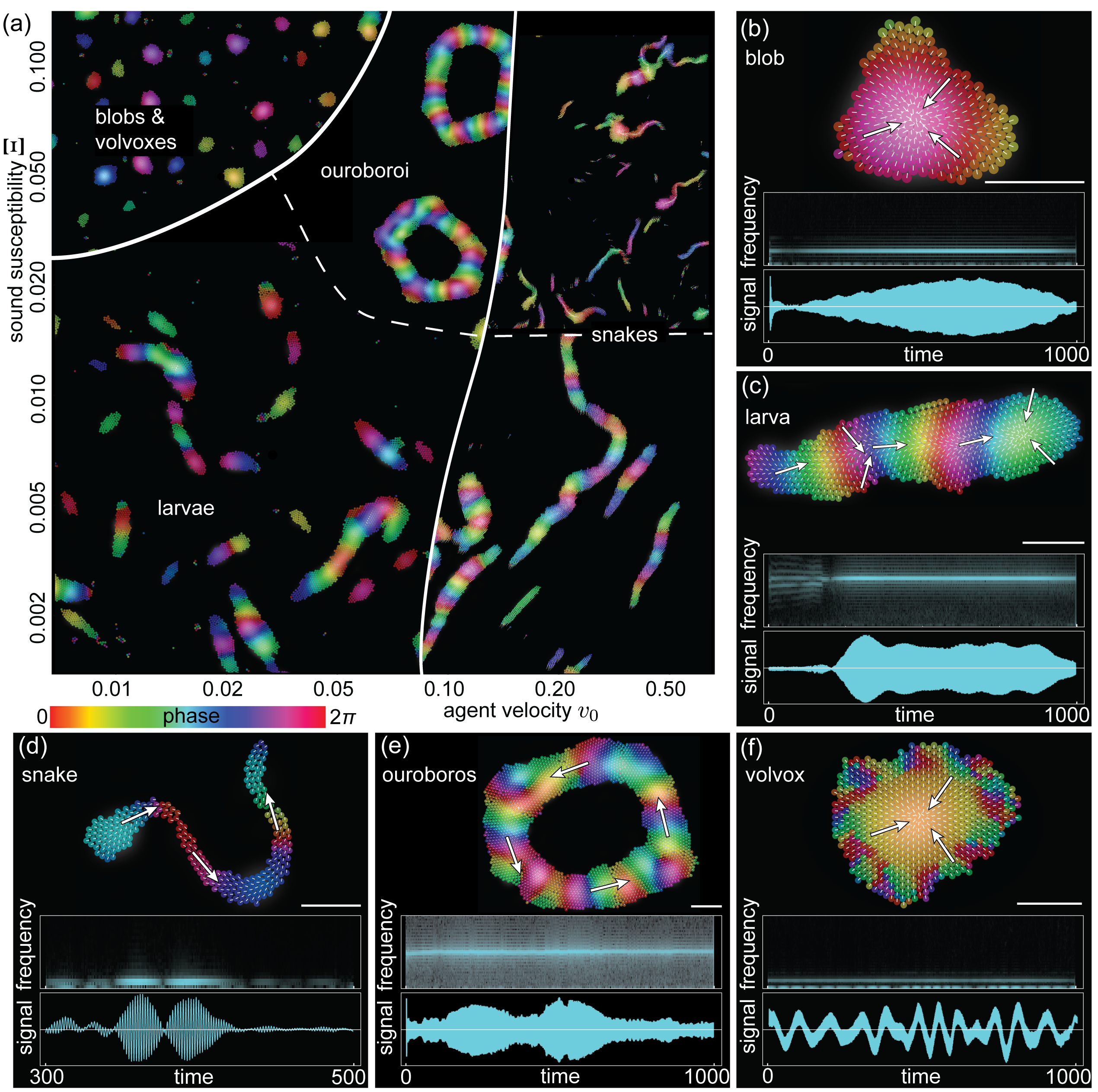}
\caption{\label{fig:pd}\textbf{Diversity of collective states.} (a) Phase diagram relating the most distinct collective states vs.\@ agent's self-propulsion velocity $v_0$  and their susceptibility to align with acoustic signals $\Xi$. Representative solutions of the agent-based model, Eq.~(\ref{eq:agents}), are shown for the five qualitatively distinct phenotypes, with the color code indicating the agents' oscillatory phase. (b-f) Representative solutions with their acoustic signatures given by frequency distribution and the acoustic signal amplitude measured in the vicinity of the collective solutions. Agent color indicates its oscillator phase, white arrows highlight the local average agent orientation and white scale bars signify a length of $5$ units.
(b) Localized blob with central polar defect and target-wave pattern. 
(c) Larva. A polar wave-emitting defect is located in its head. (d) Rapidly moving snake; no internal polar defect. Phase waves propagate along its body. (e) Ouroborus, a closed larva-like ring structure. (f) Localized volvox with a synchronized center decorated by outer circular traveling phase waves or decoherent outer layers. Below each image (b-f) is the solution's acoustic signature: spectrogram as obtained from short-time discrete Fourier transform depicted over a frequency range $0-8$ Hz (uncoupled agent's free frequency $\sim 0.07\,\text{Hz}$) and the normalized acoustic signal spatially adjacent to the respective solution. Parameters as given in Table \ref{tab:partPars}. See Supplementary Videos 2-6 for exemplary dynamics of the distinct states.}
\end{figure*}

In Fig.\@ \ref{fig:pd}, we present the predominant collective solutions as a function of the agent's velocity $v_0$ and acoustic susceptibility $\Xi$ (Fig.\@ \ref{fig:pd}(a)) and describe their phenomenology as well as their acoustic emission in terms of signal amplitudes and frequencies as measured at a nearby fixed microphone position (Fig.~\ref{fig:pd}(b-f)).
The interplay between the persistent polar motion of agents and the adaptation of this motion in response to acoustic signals are two key parameters in our model. However, there are other parameters that can affect the observed collective behavior. In appendix \ref{app:dependence}, we discuss the role of the polar alignment $\Gamma$, the acoustic coupling $\lambda$, the noise strength $\xi$, and the nonlinear frequency coupling $b$ in shaping the exhibited collective dynamics.

The simplest self-organized state is a polar aggregate (\textit{blob}) in which swarmers are aligned towards a central pacemaker (white arrows, Fig.\@ \ref{fig:pd}(b)). 
Since individual agents are oriented toward its center, they create an enclosed polar defect. 
Blob solutions predominantly occur for comparably small agent velocities and large sound susceptibilities; see Fig.~\ref{fig:pd}(a).
They emerge from a homogeneous distribution of swarmers through the initial formation of small aggregates (Supplementary Video 1). 
Then, the aggregating swarmers become tightly packed and synchronized (similar agent colors in Fig.~\ref{fig:pd}(b), color-code indicates the oscillatory phase of agents). 
Due to the alignment of agents towards larger signaling amplitudes, an almost circular region of synchronized agents develops (Fig.~\ref{fig:pd}(b)). The entire blob contributes as a large collective source to the acoustic field and emits, almost isotropically, concentric sound waves into its surroundings.
The aggregation, attraction, and mutual activation generate higher oscillation frequencies (frequency panel, Fig~\ref{fig:pd}(b), Supplementary Video 2) and amplitudes (signal panel) and, in turn, increasing sound intensities at the aggregate's center. This self-amplifying effect results in an emergent pacemaker---a region of agents with increased frequency, leading the phase dynamics---located in the blob's center. As a consequence of the increased collective sound amplitudes, the blob attracts an ever growing number of nearby swarmers. In particular, we observe a dominant oscillatory mode in the blob that is at approximately $1.8$~Hz; around $25$-fold the frequency of the uncoupled individual agents' baseline (frequency panel in Fig.\@ \ref{fig:pd}(b)). Additionally, a secondary, incoherent frequency mode at much lower frequencies is present. It represents the outer agents in the collective that are not fully synchronized with the center and experience weaker acoustic inputs. From the normalized signal amplitudes of the solution (lower panel in Fig.\@ \ref{fig:pd}(b)) we see that amplitudes at a fixed microphone position are modulated over long time scales, representing the slow drifting motion of the blob solution. As a rather synchronized solution, blobs create a standing wave field around them and the modulation of the amplitude reflects the drift of the maximum position of the standing wave, see Supplementary Video 2.

Another example of a self-organized, symmetry-broken state is an elongated, slowly migrating aggregate that we term a \textit{larva} (Fig.\@ \ref{fig:pd}(c), Supplementary Video 3). 
Here, a broken symmetry in the position of the enclosed polar defect or defect line—resulting from an asymmetric aggregation process or the merging of two blobs—may cause the larva to move slowly but persistently. Larvae are observed for similarly low agent velocities as blobs but typically at weaker acoustic susceptibilities (Fig.\@ \ref{fig:pd}(a)). 
In this parameter regime, agents experience weaker torques toward the phase leader (e.g., pacemaker) with the highest sound amplitudes and the co-localized polar defect. 
Thus, asymmetries around that polar defect may emerge more naturally.
As apparent from numerical simulations, the asymmetrically confined pacemaker in the larva emits phase waves, aligning all other swarmers in the aggregate towards it.
Given these characteristics, the larva acts as a motile community that can absorb individual clusters and reintegrate them into its structure; see Supplementary Video 1. 
Investigating its acoustic fingerprint, we observe that the aggregate displays high collective frequencies (Fig.\@ \ref{fig:pd}(c), frequency panel). 
For the chosen parameters, these frequencies are approximately $80$ times higher than the frequency of uncoupled individual agents.
Such a significant frequency increase requires a comparably large amplitude of the collective acoustic field. Hence, we propose that phase differences between neighboring agents within the larva align with the acoustic wavelength, enabling all contributing agents to cooperatively reinforce one another. As a result, signal amplitudes increase and mutual phase velocities accelerate. Consistently, the acoustic amplitudes received from the larva (Fig.~\ref{fig:pd}(c), signal panel) rise considerably as it approaches the microphone position. As it passes by, the sound amplitudes remain relatively stable, suggesting a nearly uniform coupling across the entire collective. However, a slight amplification occurs when the head of the structure is closest to the acoustic detector ($t\approx 350$). This highlights a key characteristic of the larva as its head houses the pacemaker, the region with the highest frequency and amplitude (Fig.\@ \ref{fig:pd}(c), signal panel), see Supplementary Video 3.
Compared to the blob solution discussed above, the larva completely passed the microphone position with its entire length in the time-frame shown, whereas the blob solution just slightly drifted into a minimum of the standing acoustic wave. 
This distinction highlights the persistent, directed motion of the larva solution and represents another significant difference in the solutions' acoustic signatures.

Rapidly moving \textit{snakes} is yet another example of states with collective functionality (Fig.\@ \ref{fig:pd}(d), Supplementary Video 4). 
They occur at large agent velocities where the directed propagation of agents gives rise to the collective, snake-like motion of these structures.
Snakes lack an internal pacemaker such that phase waves of the oscillation typically propagate from head to tail.
The resulting acoustic field aligns all agents in a common direction along the emerging center line as highlighted by the white arrows in Fig.\@ \ref{fig:pd}(d).
Somewhat similar collective snake-like states have been observed in active matter with vision cones~\cite{negi2024collective}.  
In our study, the latter structures emerge via spontaneous symmetry breaking from the acoustic interactions and the self-organized information propagation through the phase waves.
The mutual alignment of all the agents within the snake results in a collective propagation velocity of snakes that is comparable to the single swarmer speed $v_0$. 
Unlike individual agents, snakes exhibit significantly higher persistence of motion due to the coordinated alignment of neighboring agents. 
This enhanced coordination leads to rapid collective propagation, which is reflected in the acoustic signal detected by a stationary microphone. The signal shows a brief duration and rapid modulation as the snake passes by.
Additionally, because snakes lack an internal pacemaker and high self-sustained acoustic field amplitudes, the oscillation frequencies of the agents in the snakes are relatively low, comparable to those of uncoupled agents. This is evident from the signal and frequency panels shown in Fig.\@ \ref{fig:pd}(d).

When increasing the sound susceptibility $\Xi$, larvae can transform into rotating ring-like entities (top center in Fig.\@ \ref{fig:pd}(a)). 
Since this process involves the larvae curling into a circular shape and metaphorically ``eating" their tails to form a continuous loop, we refer to these entities as \textit{ouroboroi}, inspired by an ancient symbol for eternal cyclic renewal (Fig.\@ \ref{fig:pd}(e), Supplementary Video 5). 
These ouroboroi display periodic phase waves that propagate through the entire structure, as can be seen from the color-coded phases.
Typically, these phase waves propagate in the direction opposite to the motion of agents, indicated by the white arrows. 
Since ouroboroi are essentially ``closed-larvae" states, where rotation is limited due to mutual blocking of the agents, their acoustic signatures closely resemble those of larvae solutions.
However, as ouroboroi are mainly localized at a given position, the emitted acoustic signal does not decay as strongly as for the passing larvae; see bottom panels in Fig.\@ \ref{fig:pd}(e).
Ouroboroi are reminiscent of closed ring-like structures observed in systems of large-scale chemical attraction \cite{Romanczuk2008Beyond, ziepke2022multi}, where ring formation is mediated by an interplay between attraction to an aggregation center and the persistent motion of agents. Similarly, as seen in the milling motion of agents with vision-cone alignment \cite{barberis2016large, negi2024collective}, the phase-wave synchronization dynamics induced by acoustic interactions lead to alignment with the agents ahead and allow for ring formation.
Finally, if the structure size of blobs exceeds the range over which constituting agents can acoustically synchronize, the aggregates can become decorated by outer layers of agents with traveling, metachronal waves or desynchronized oscillations (\textit{volvoxes}) (Fig.\@ \ref{fig:pd}(f), Supplementary Video 6). 
Therefore, heterogeneous phase patterns or decorrelated oscillations surround the central phase-synchronized region. This situation is reminiscent of the ``chimera states'' occurring in coupled oscillator systems~\cite{abrams2004chimera}.
The coexistence of synchronized and desynchronized swarmers is also reflected in the agents' frequency distribution, where we observe a constant part with increased frequencies, about 10-fold that of free agents, and an irregular fraction of agents with lower frequencies (frequency panel in Fig.\@ \ref{fig:pd}(f)). 
In the example shown in Fig.\@ \ref{fig:pd}(f), featuring metachronal waves, the emitted waves interfere with the center synchronized oscillations and lead to a strong modulation of the acoustic signal.

Altogether, we observe multiple emergent collective states in the system of acoustically coupled swarmers. Blobs and volvoxes arise predominantly for small agent velocities and when the sound-mediated attraction is strong (Fig.\@ \ref{fig:pd}(a)). For weaker acoustic susceptibility/larger agent velocities, aggregates can become asymmetric, leading to the formation of larvae or ouroboroi solutions. 
Increasing the agent velocities further, the collectives lose any internal polar defects, and snake solutions become predominant. The collective solutions not only feature phenomenological differences in terms of their polar defect localization or collective propagation velocity but also exhibit distinct acoustic signatures. The acoustic coupling induces different collective frequency distributions with varying mean and spread and distinguishable dynamics of acoustic amplitudes. 
Thereby, the system of acoustically interacting agents gives rise to cognitive flock configurations through spontaneous symmetry breaking and self-organized frequency distribution of agents via acoustic coupling. In previous studies of active matter systems, similar types of collective states were observed only upon explicitly imposing a symmetry-breaking vision cone for mutual interactions \cite{barberis2016large}.

\subsubsection{Transitions between the collective states}\label{sec:pd_quant}
How can one quantify the transitions between these various self-organized collective states? 
As we have seen, the primary phenotypes (blobs, larvae, and snakes) qualitatively differ in their collective self-propulsion behavior. Blobs, characterized by a central polar defect, remain relatively localized, whereas larvae, with asymmetrically positioned polar defects, exhibit collective motion at intermediate velocities. In contrast, snakes that lack an internal polar defect, move collectively at speeds approaching those of free, individual agents. These distinct phenotypical behaviors enable a quantitative analysis of the observed phases in which different collective states dominate the system's long-term dynamics.
While there is a large set of possibilities for quantification, we opted for characterizing the degree of polar order of the collective states in terms of a \textit{cluster polar order parameter} defined as
\begin{align}\label{eq:order_pol}
    \Psi_\text{pol}=\frac{1}{v_0\,N}\sum_\text{clusters}\bigg\lvert\sum_{i\in\text{cluster}}\vec{v}_i\bigg\rvert\,,
\end{align}
with total agent number $N$.
This quantity characterizes the averaged polar order within clusters, weighted by the number of agents constituting them.
Additionally, we examine the overall clustering behavior observed in numerical simulations of the system using the order parameter \cite{Zhao2021Phases}
\begin{align}\label{eq:clustering}
    \Psi_\text{clust}=\frac{1}{N}\sqrt{\sum_{i\in\text{clusters}}N_i^2}\,.
\end{align}
This \textit{clustering measure} is given as the normalized $L_2$ norm of cluster sizes, each identified by the number of contributing agents $N_i$. We define individual clusters as groups of agents in which each member is connected to at least one other agent within the cluster by a pairwise distance less than $2.5$ agent radii $r_p$.
The quantity $\Psi_\text{clust}$ gives a measure for the formation of large clusters and allows us to distinguish between dispersed agents, small, separated blobs, large aggregated larvae, and smaller collectives of snakes.

\begin{figure}[!tb]
    \centering
    \includegraphics[width=\linewidth]{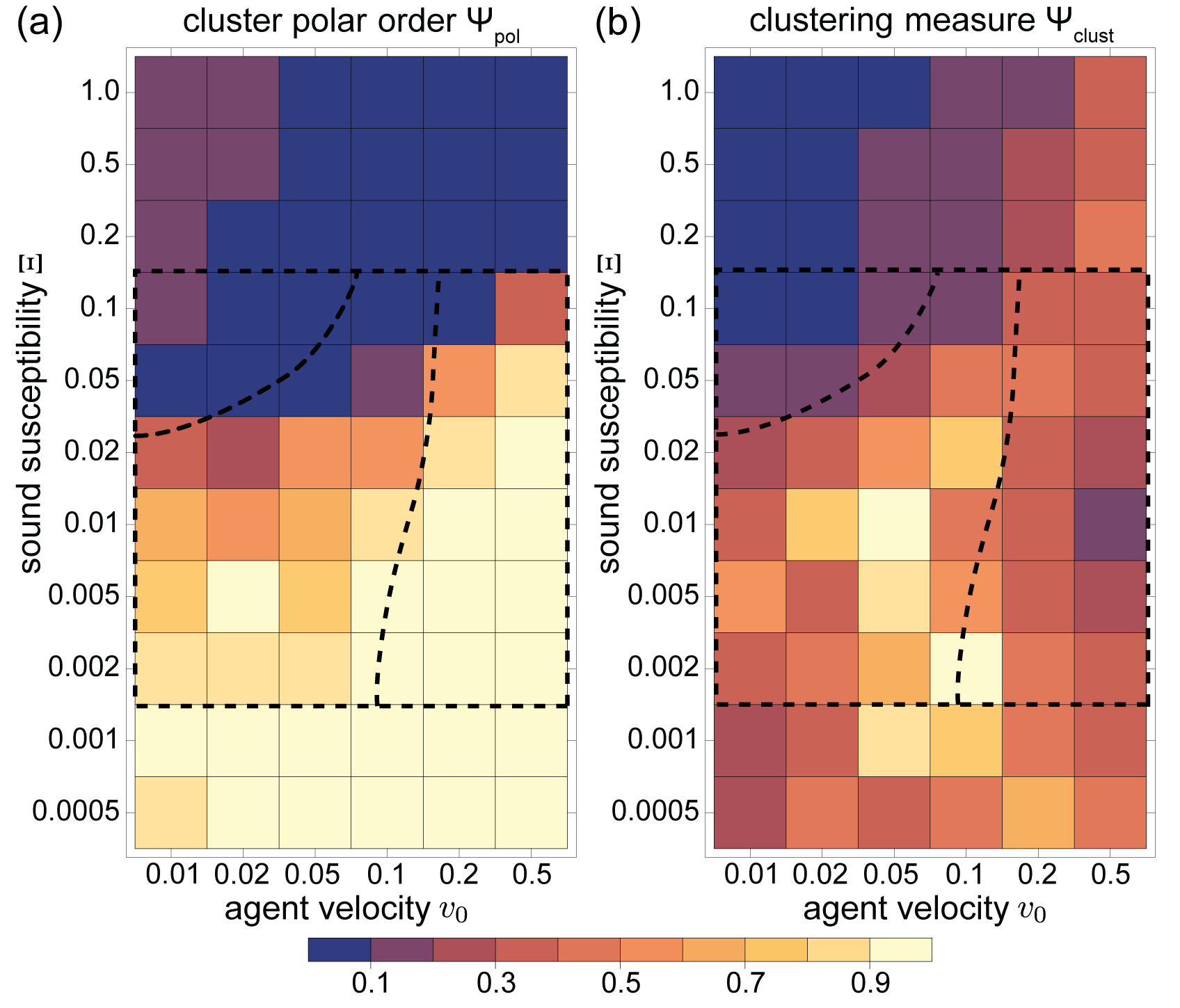}
    \caption{\textbf{Quantification of phenotypical behavior.} The transitions between predominantly occurring phenotypical states (dashed lines, Fig.~\ref{fig:pd}) correspond to the quantitative analysis of the numerical solutions: (a) Average cluster polar order parameter $\Psi_\text{pol}$, Eq.~\eqref{eq:order_pol} and (b) clustering measure $\Psi_\text{clust}$, Eq.~\eqref{eq:clustering}. Corresponding final states of numerical simulations underlying the phase diagram are shown in appendix Fig.~\ref{app_fig:quant_pd}.}
    \label{fig:quant_pd}
\end{figure}

Evaluating the cluster polar order $\Psi_\text{pol}$ and the clustering measure $\Psi_\text{clust}$ for numerical simulations of the agent-based model equations, Eqs.~(\ref{eq:agents}), yields the quantitative phase diagram, Fig.~\ref{fig:quant_pd}. We observe that the average polar order within clusters differentiates the top part with high acoustic susceptibilities from the lower part of the parameter space, Fig.~\ref{fig:quant_pd}(a). This is because, for large values of $\Xi$, agents are strongly attracted towards local aggregation centers where they form aster-like polar defects within blob structures. Moreover, a trend can be observed in the cluster’s polar order, distinguishing the slightly higher polar order of snake-like solutions ($\Psi_\text{pol} > 0.9$) from that of larvae ($\Psi_\text{pol} < 0.9$). This is because the larva structures contain a polar defect and, consequently, agents that are oriented against the net motion of the collective.\\
Considering the clustering measure $\Psi_\text{clust}$, Eq.~(\ref{eq:clustering}), in Fig.~\ref{fig:quant_pd}(b), we observe the lowest values in the regime of blob solutions as, there, various blobs compete as local aggregation centers and their merging and the formation of large blob solutions which would correspond to large cluster measures is slow. In the parameter regime where snakes are dominant, the clustering measure is elevated to intermediate values, $0.2\lesssim \Psi_\text{clust}\lesssim 0.6$. Large larvae represent the attractors that aggregate most agents in the phase diagram, Fig.~\ref{fig:quant_pd}. Accordingly, they can be identified by high clustering measures ($\Psi_\text{clust}\gtrsim 0.3$) and intermediate values of cluster polar order $\Psi_\text{pol}$.
Altogether, the measures $\Psi_\text{pol}$ and $\Psi_\text{clust}$ enable the quantification of the observed collective behavior in the system, allowing for the distinction of regions in parameter space that exhibit different predominant phenotypical collective solutions. However, numerical simulations also reveal (Fig.~\ref{app_fig:quant_pd}) that the transition lines between these regions correspond to parameter ranges of metastability, where qualitatively different phenotypical states can coexist. The observed collective behavior to some extend depends on the system’s initial configuration. Furthermore, variations in the cluster polar order parameter $\Psi_\text{pol}$ and the clustering measure $\Psi_\text{clust}$ within a given dominant phenotypical region in Fig.~\ref{fig:quant_pd} suggest that the expression of the distinct phenotypes is not uniform but varies across parameter space, see App.~\ref{app:pd} for details.

\subsection{Continuum equations}

\begin{figure*}
\centering
\includegraphics[width=\linewidth]{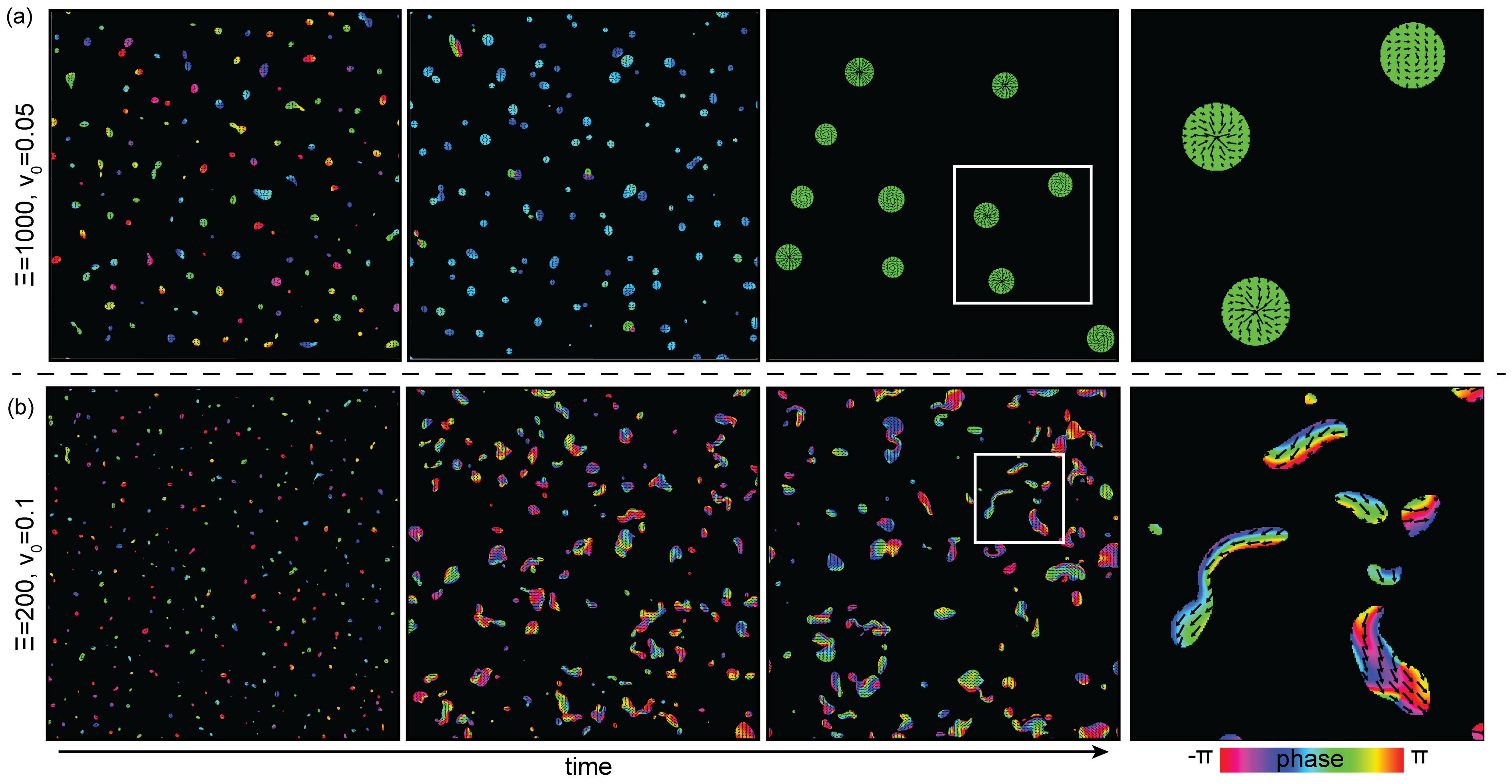}
\caption{\textbf{Collective states in the continuum model.} Temporal evolution of the continuum field representation of acoustic active matter, Eqs.\@ \eqref{eq:field}. Different types of solutions emerge from a disordered homogeneous state. (a) Prevalence of localized blobs for $v_0=0.05$ as shown in the zoom-in on the right side panel. (b) Snakes emerge upon initial blob formation and partially aggregate into larger blobs for $v_0=0.1$. Snake solutions in the continuum model are highlighted in the zoom-in (rightmost panel). The color indicates local oscillatory phases for densities $\rho>1$. Arrows represent the polar orientation $\bm{p}(\bm{r},t)$. The system size is set to $L=200$, and zoom-ins detail the highlighted domains. Remaining parameters as given in Table \ref{tab:fieldPars}.}
    \label{fig:field}
\end{figure*}

In this section, we explore the collective dynamics in the continuum model, given by Eqs.~(\ref{eq:field}). 
We begin by comparing the emerging structures with those found in the agent-based model. 
Then, we leverage the continuum field equations to study the behavior of acoustic active matter at large time and length scales, a regime beyond the reach of the agent-based simulations.

As for the agent-based description, we observe a rich phenomenology due to the acoustic coupling of swarmers (Fig.~\ref{fig:field}, Supplementary Videos 7 and 8). 
Upon formation of small clusters, we observe coarsening behavior with the swarmers merging into larger aggregates. 
In particular, as in the agent-based model, the system dynamics strongly depends on the susceptibility parameter $\Xi$ and the swarmer velocity $v_0$.
Namely, at low self-propulsion velocities, ${v_0=0.05}$, we see the formation of multiple small blobs which subsequently coarsen to fewer larger ones (Fig.\@ \ref{fig:field}(a)); see zoom-in in the rightmost panel. 
Here, an aster-like polar orientation of the swarmer matter towards a central defect is prevalent. 
In addition to what has been observed for the discrete system, the field equations also exhibit vortex-type blob solutions with chiral motion around the central defect. 
At larger velocities, ${v_0=0.1}$, and weaker signal susceptibility, $\Xi=200$, we observe a prevalence of snake-type structures (Fig.\@ \ref{fig:field}(b)); see zoom-in for details. 
As for the agent-based case, these structures are highly motile and free of internal pace-making polar defects. They typically collide and aggregate into larger structures, but occasionally also split such that new smaller snakes emerge.
The full temporal dynamics for the two cases are shown in Supplementary Videos 7 and 8, respectively.

\begin{figure}[!t]
\center
\includegraphics[width=\linewidth]{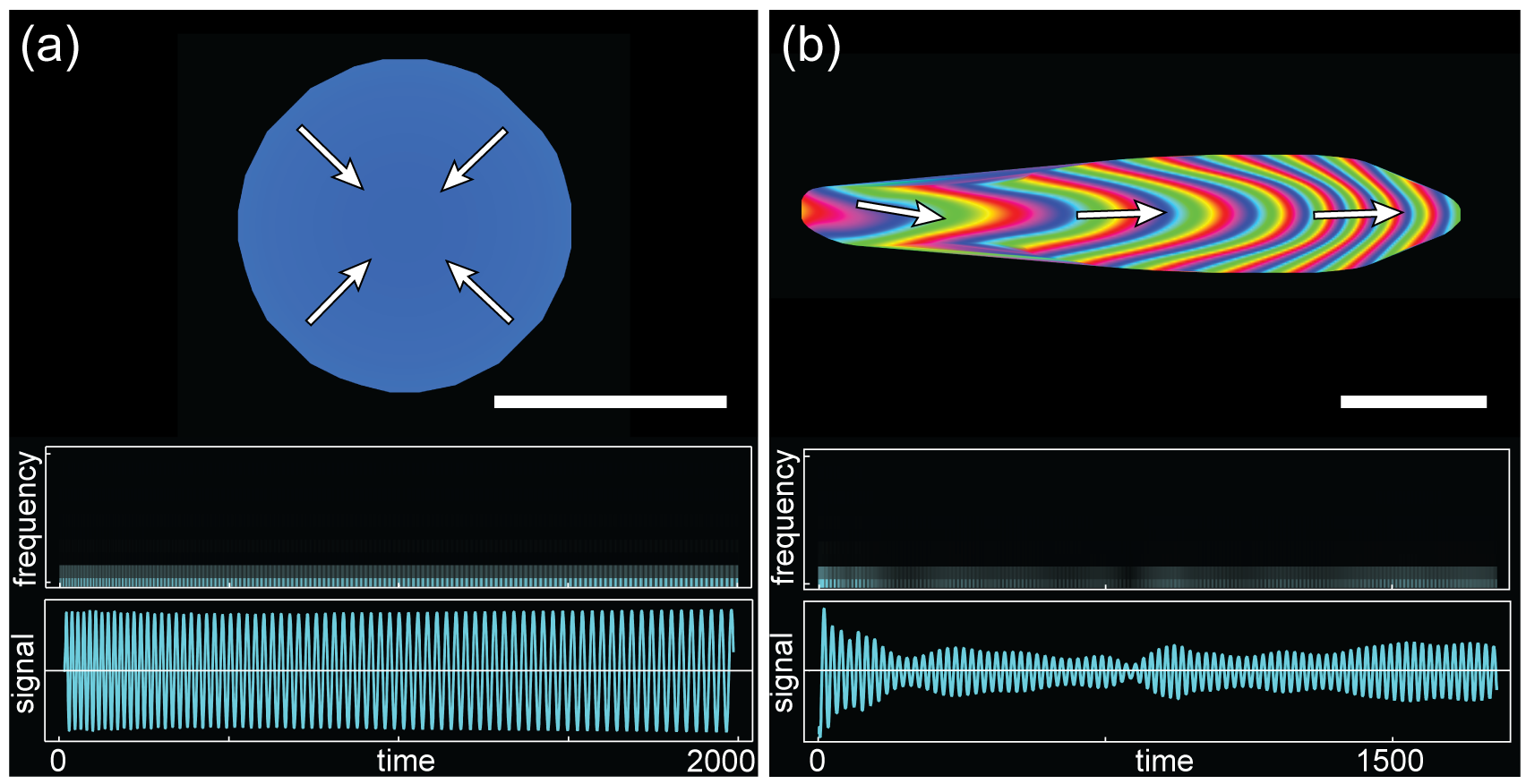}
\caption{\textbf{Acoustic signatures of aggregates in continuous field description.} (a) Representation of a blob solution. Densities $\rho>1$ are color-coded by their oscillatory phase (see color bar, Fig.~\ref{fig:field}). The swarmers are oriented towards a central polar defect; white arrows indicate the average polar direction. (b) Snake-like solution. Phase waves propagate from head to tail through the structure. The resulting average polar orientation (white arrows) of agents leads to a net motion of the entire aggregate. White bars represent a length of $10$ units and spectrograms show a frequency range from $0-4\,\text{Hz}$.}
\label{fig:field_sound}
\end{figure}

In summary, the continuum model captures two distinct collective phenomena also observed in the agent-based model: blobs (Fig.~\ref{fig:field}(a)) and snake-like aggregates (Fig.~\ref{fig:field}(b)).
Similar to the agent-based model, blobs in the continuum model also exhibit a high degree of synchronization, as evidenced by the uniform phase color in Fig.~\ref{fig:field_sound}(a).
As swarmers are aligned to the blob's center, the collective has an almost vanishing net motion. 
Combined with the stable synchronization of the oscillatory matter within, the amplitude of the emitted signal shows only slight modulation (see signal panel).
Oscillation frequencies of the swarmer matter are slightly increased compared to the uncoupled case, i.e., without input from the collective acoustic field. 
In contrast, snake solutions (Fig.~\ref{fig:field_sound}(b)) are characterized by stronger fluctuations in their acoustic signals (signal panel in Fig.~\ref{fig:field_sound}(b)). 
Compared to blob solutions, the snakes exhibit slightly lower oscillation frequencies (frequency panel) due to the lack of a phase leader with increased collective stimulus.

Notably, we do not observe larva-type structures in the continuum model. 
We rationalize this as follows:
In the agent-based model, the net motion of larvae results from short-range repulsion between individual agents and the asymmetrical pushing of agents around the enclosed defect. 
In contrast, in the field theory, the finite size of the swarmers is represented by gradients in the swarmer density, which show only slight variation within the aggregates. 
Thus, swarmers are mainly advected in the direction of their self-propulsion ---towards the $+1$ polar defect--- implying a lack in persistent net motion of defect-enclosing structures. 
Since the net motion of larvae arises from high-density effects, it is unlikely to be fully captured by the gradient expansion that underlies the continuum model. However, exploring an even broader range of model parameters, such as for the self-advection term $\chi$, or incorporating additional effective terms in the dynamics that account for the mechanisms driving larva-like motion could further improve the correspondence between the agent-based simulations, Eqs.~(\ref{eq:agents}), and the continuous field equations, Eqs.~(\ref{eq:field}) 

\begin{figure*}[!ht]
\centering
\includegraphics[width=\linewidth]{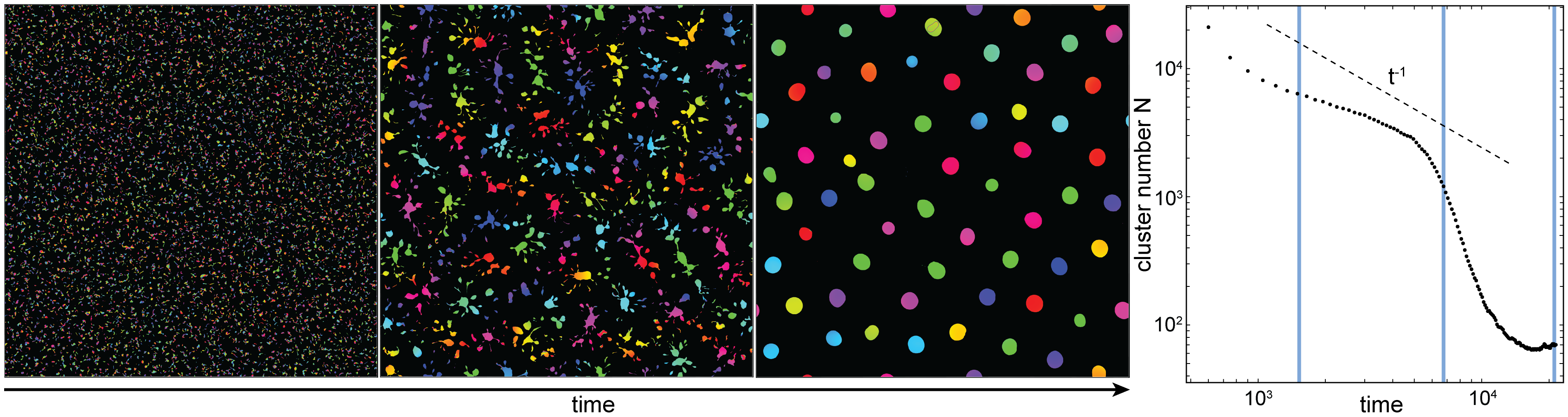}
\caption{\textbf{Large scale dynamics of acoustic active matter.} Upon the emergence of clusters, the aggregates coarsen mediated by acoustic interactions (rightmost panel). Vertical blue lines indicate the time points of the snapshots (left). Densities are shown for $\rho(\bm{r},t)>1$, and color code represents oscillatory phase $a(\bm{r},t)$ (color scale as in Fig.~\ref{fig:field}). System length ${L = 2000}$. Parameters as stated in Table~\ref{tab:fieldPars}.}
\label{fig:field_coarsen}
\end{figure*}

However, having established the overall phenomenological relation between the emergent behavior in the agent-based model and the continuous field representation, we next discuss the large-scale dynamics (Fig.~\ref{fig:field_coarsen}).
We observe that the acoustic coupling not only enables the formation of aggregates in the system but it also yields coarsening dynamics that is much faster than typical Cahn-Hilliard-like Ostwald ripening. 
Small aggregates interact via the emission of acoustic waves and synchronize their collective oscillations with neighboring aggregates through the acoustic soundscape. 
This leads to the formation of large-scale synchronization patterns that overspan multiple clusters (phase colors in snapshot $t=2000$, Fig.~\ref{fig:field_coarsen}).

This large-scale synchronization between clusters promotes their merging and speeds up aggregation into fewer larger structures. 
Compared to local diffusive interface-mediated ripening, which would induce coarsening dynamics for which the cluster number scales as ${N \sim t^{-1}}$, the acoustic synchronization yields an accelerated coarsening. 
Reminiscent of the defect coarsening discussed for the non-locally (acoustically) coupled complex Ginzburg-Landau subsystem (App.~\ref{app:cgle}), the cluster number saturates for large times and coarsening is halted at a particular structure distance. 
As for the defect coarsening in the non-locally coupled complex Ginzburg-Landau equation, this inter-cluster distance may be selected by the wavelength of the established acoustic wave field.

Overall, the continuum field equations are a complementary way of assessing the relevance of acoustic interaction in active matter. 
They reveal a phenomenology that closely parallels the agent-based model, which enables us to draw broader conclusions about the role of acoustic coupling in the large-scale dynamics of the system. 
Specifically, we find that acoustic waves facilitate nonlocal phase synchronization and mutual attraction between aggregates, which in turn regulate the length scales of emergent structures and their interactions. 
By mediating these long-range effects, acoustic coupling plays a crucial role in shaping the collective behavior and spatial organization of the system.

The field equations examined in this study provide valuable insights into the large-scale dynamics of acoustic active matter. 
Beyond this preliminary investigation, they open up a wide range of potential applications, including wave-coupled swarming systems in heterogeneous media and general acoustically coupled active media, along with various others.
While the field equations qualitatively capture some key solutions including their acoustic signatures, they fail to represent others such as the larva solutions. 
This limitation arises because the continuous model, based on a gradient expansion, does not account properly for high density effects that drive the net motion of the larvae.
In the following, we aim to explore potential applications of the emergent collective solutions which involve a tractable number of agents. 
To better incorporate potential small-number effects and gain more accurate insights, we turn again to the agent-based model.

\section{Collective functions}

The wide spectrum of self-assembled states (Fig.\@ \ref{fig:pd}) provides the opportunity to tailor these states for specific functions.
As we have demonstrated, acoustic active matter generates self-organized collective states with distinct acoustic signatures. 
By leveraging this self-organized behavior, we next show how to harness the emergent collective functionality for a range of practical applications.

In general, individual agents in active matter systems are small in comparison to the collective structures and, as in our case, have limited processing capabilities. 
Furthermore, they can only access local information. 
Therefore, potential applications rely on cooperation and require a collectively synchronized behavior.
This synchronization can be achieved through acoustic interactions.
Importantly, since the systems' dynamics is relying on self-organization it offers inherent robustness of the emergent collective states against perturbations such as environmental changes or failure of individual swarmers. 
For collectives of identical swarmers, self-organization yields a behavioral identity ---that is, a kind of phenotypic state--- by which agents develop different oscillation frequencies and exhibit self-organized motion due to their acoustic coupling. As this identity is emergent, individual swarmers can be seamlessly replaced without compromising the functionality of the group.

With regard to potential applications, emergent collectives must possess the ability to identify where their action is needed, move accordingly, and assess the effectiveness of their action. This crucially requires adaptability to their environment and a dynamic, coordinated response to external cues.
Additionally, although self-organization drives these emergent structures, external control may still be necessary to harness and direct the collective’s functionality. 
This raises key questions:  how can collectives adapt to changing conditions, and how can they be externally controlled?
Acoustic signaling, in principle, provides a mechanism that enables both these crucial elements. 
Environmental changes can be detected through the reflection of collectively emitted acoustic signals from passive objects. 
A collective response to external stimuli is achieved by modulating the oscillatory behavior of the swarmers, where changes of the oscillations and the mutual interactions between swarmers enables the coordination of collective behavior.
In the following, we will address these questions by investigating the self-organization capabilities that emerge at the collective level, including the collective perception of the structures and their response to external control inputs.

\subsection{Collective sensing}\label{sec:function_sensing}

\begin{figure*}[!ht]
\includegraphics[width=\textwidth]{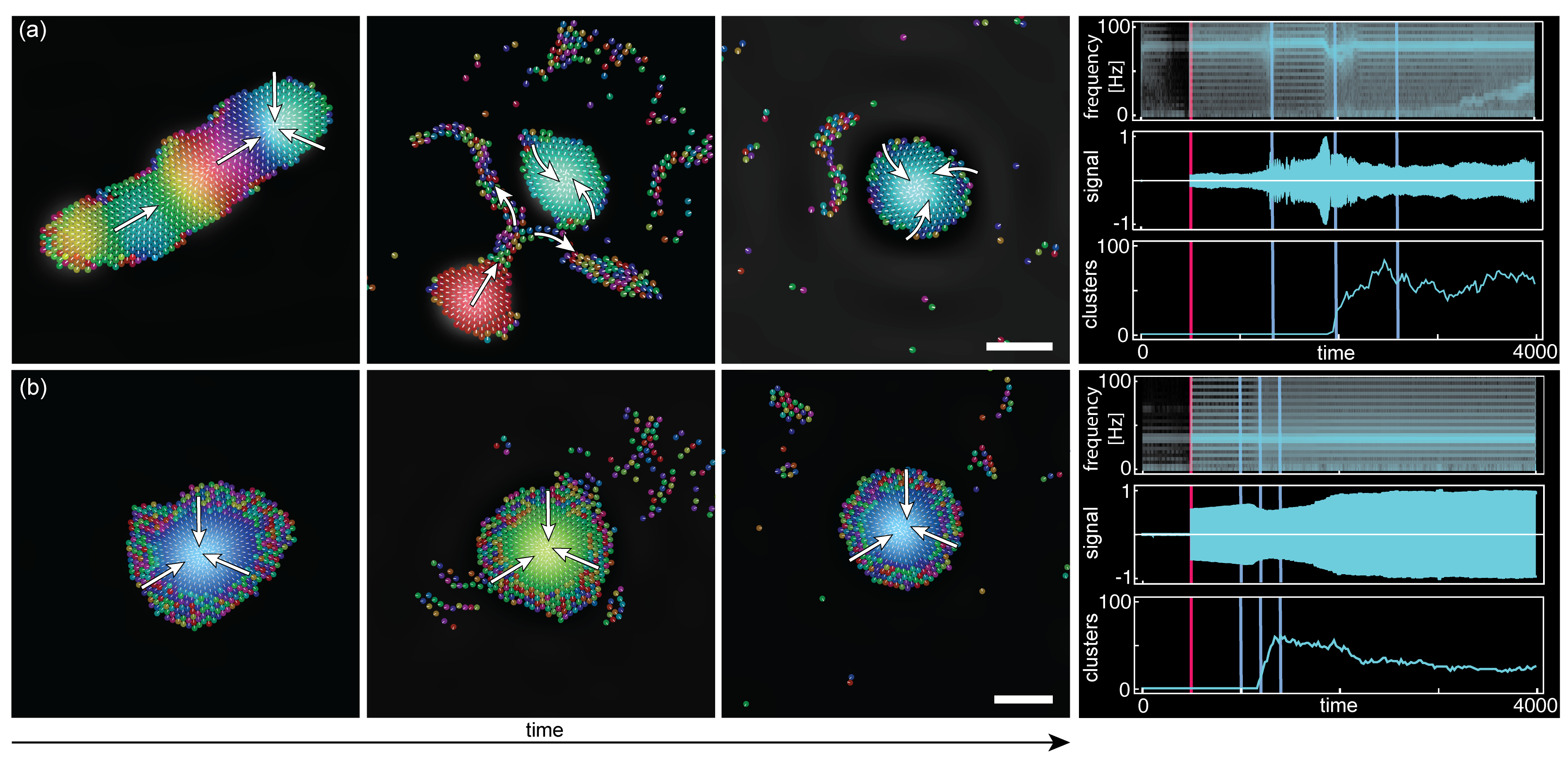}
\caption{\textbf{Collective sensing of approaching objects.} (a) A sequence of snapshots illustrating the phenotype change of a traveling collective solution (larva) in response to an approaching object (``intruder'' / ``threat'') above the habitat; the acoustic signals are reflected at the external object. The temporal evolution of sound frequencies, amplitudes, and the number of individual clusters in the entire simulation domain $L=50$ are displayed in the rightmost panel. The vertical red line marks the instance when an intruder is introduced. Blue lines indicate times corresponding to the snapshots. The oscillation amplitude is measured in dimensionless units of Eq.\@ (\ref{we}). 
 (b) A sequence of snapshots illustrating the response of a localized collective solution (blob) on an approaching intruder. Disks represent individual agents, and the RGB color map shows the relative phases; see Fig.~\ref{fig:pd}. White arrows indicate the agent's polar orientation and white bars represent a length of $5$ units.}
 \label{fig:sensing}
\end{figure*}

A key capability of systems that use (acoustic) waves for communication is their ability to emit and detect signals. 
Beyond transmitting signals between agents, these active systems can also acquire information about their surroundings through the reflections of the acoustic waves of various objects.
However, this requires a collective response of the system, since a single agent on its own cannot generate a significant response due to its relatively weak emission amplitudes. 
When agents are organized in collective states, as discussed above, they are able to synchronize and emit acoustic waves in unison, substantially boosting the amplitudes of the waves. 
This coordination results in emergent cooperative sensing, where the enhanced signal strength and coordinated emissions enable more effective detection and interpretation of signals reflected from invading objects.

As an illustration of such a sense-and-response capability, we consider the response of a propagating larva and a localized blob to an invading object (Fig.\@ \ref{fig:sensing}).
As the external object descends toward the habitat, the acoustic waves emitted by the collective state get reflected at the object. This, in turn, is perceived by the collective as it results in a change in their oscillatory states.
In particular, we model the reflective object as a plane above the agents' habitat that approaches from the top with a distance $d_z$ following the protocol
\begin{align}
    d_z(t)=4\,d_\text{final}\left(1-\frac{t-t_0}{2000}\right)\,,\,\, \mbox{for }t_0<t<2000\,,
\end{align}
and remains at the value $d_z=d_\text{final}$ for larger times.
We insert the object after an initial time $t_0=500$ at height $4\,d_\text{final}$ and let it approach a distance of $d_\text{final}$ after $1500$ units of time.
For simplicity, we assume that the waves reflected at the object can be approximately described as quasi-planar and with a homogeneous phase. 
This approximation is valid for objects with sufficiently large distance (height above the habitat). 
Then, the additional reflective input to Eq.\@ (\ref{we}) is given by 
\begin{eqnarray}
    u_{\text{refl}}(\bm{r},t)=A_{\text{refl}}\,\langle a_l\rangle_N\,e^{2i\omega d_z(t)/c}\,.
\end{eqnarray}
For the assumed spherical wave solution in three-dimensional space that decays inversely proportional to the distance, the sound amplitude depends on the distance $d_z(t)$ as ${A_{\text{refl}} = 1/\left(2d_z(t)\right)}$. 
If there is a non-vanishing contribution $\langle a_l\rangle_N$ averaged over all the $N$ acoustic oscillators, the agents receive a reflected phase-shifted acoustic feedback signal. 
The strength of the reflected signal not only depends on the distance of the reflective object but is significantly determined by the degree of synchronization in the emitting collective. 
The more agents emit signals in unison, the stronger the reflected signal will be. 
Thus, the system shows a cooperative exploration of the surroundings.

In our simulations, with a final distance $d_\text{final}=50$ of the reflected object above the habitat, we observe that upon detecting the reflected waves (left panel in Fig.\@ \ref{fig:sensing}(a), left blue indicator in the right panel), a larva solution undergoes a dramatic transformation of its morphology (`metamorphoses', center), disassembling into a blob and expelling peripheral swarmers (right panel in timeseries), see Supplementary Video 9.
The ejection of agents (center blue time marker) is evidenced by the strong increase in the number of clusters in the entire simulation domain (`clusters' panel). 
Right before this event, the signal amplitudes increase significantly (`signal' panel) as more and more agents of the larva start to synchronize their oscillatory state. Apparently, at some point, local synchronizations within the larva become too strong. The larva destabilizes and starts to metamorph into an intermediate state in which its head and tail form two separate blob-like structures.
One of them further emits agents into the surroundings and eventually decays. This event corresponds to a dip in the aggregate's oscillation frequency, as emitted swarmers decouple from the center aggregate. The other developing blob remains spatially localized and shows a slight vorticity. While the swarmers within the remaining blob display synchronized oscillations, the expelled swarmers have decoherent phases. Finally, some of the dispersed agents aggregate into a secondary snake-like structure (second growing mode in `frequency' panel) which is spatially captured in a high-amplitude acoustic wave ring emitted by the blob (Fig.~\ref{fig:sensing}(a), last snapshot).

Like larvae, blobs, and volvoxes can also detect and respond to approaching objects via reflected signals (Fig.~\ref{fig:sensing}(b)); see Supplementary Video 10. 
Once the reflective object is introduced above the habitat (red marker in the right panel), the blob's acoustic signature is changed: 
The amplitude of acoustic emission increases significantly given the additional input through reflection at the intruder.
As the reflective object approaches the habitat, the volvox sheds peripheral, desynchronized agents, reducing its overall size (clusters panel). 
Thereby, the aggregate takes a more circular shape, as agents further away from the synchronized center can no longer be attracted. 
At this point, the volvox has reached its carrying capacity. 
The ejection of excess swarmers also leads to a slight reduction of the dominant oscillation frequency (frequency panel) and a dip in the signal amplitude (signal panel); see Supplementary Video 10 for the corresponding temporal evolution.

In both these cases, the collective states responded to the reflective object by altering their collective behavior. 
The observed self-organized responses to intruders, along with the resulting behavioral changes, can be linked to specific collective functions. 
For instance, the morphological transition from a larva to a blob corresponds to an externally induced localization of the aggregate. 
Similarly, the volvox's response can be seen as a self-organized, coordinated dispersion of agents in reaction to the intruder.
Since we did not specifically design the system with this functionality in mind, the observed behavior is emergent, demonstrating that minimal physical interactions between agents can give rise to a higher-order collective functionality. Again, we have observed a correspondence between the emergent behavior of the aggregates and their collective acoustic signatures. Through acoustic coupling, the agents emit distinct state information into their surroundings, suggesting a rudimentary form of inter-collective communication.

\subsection{Cooperative functionality}

\begin{figure*}
\centering
\includegraphics[width=\linewidth]{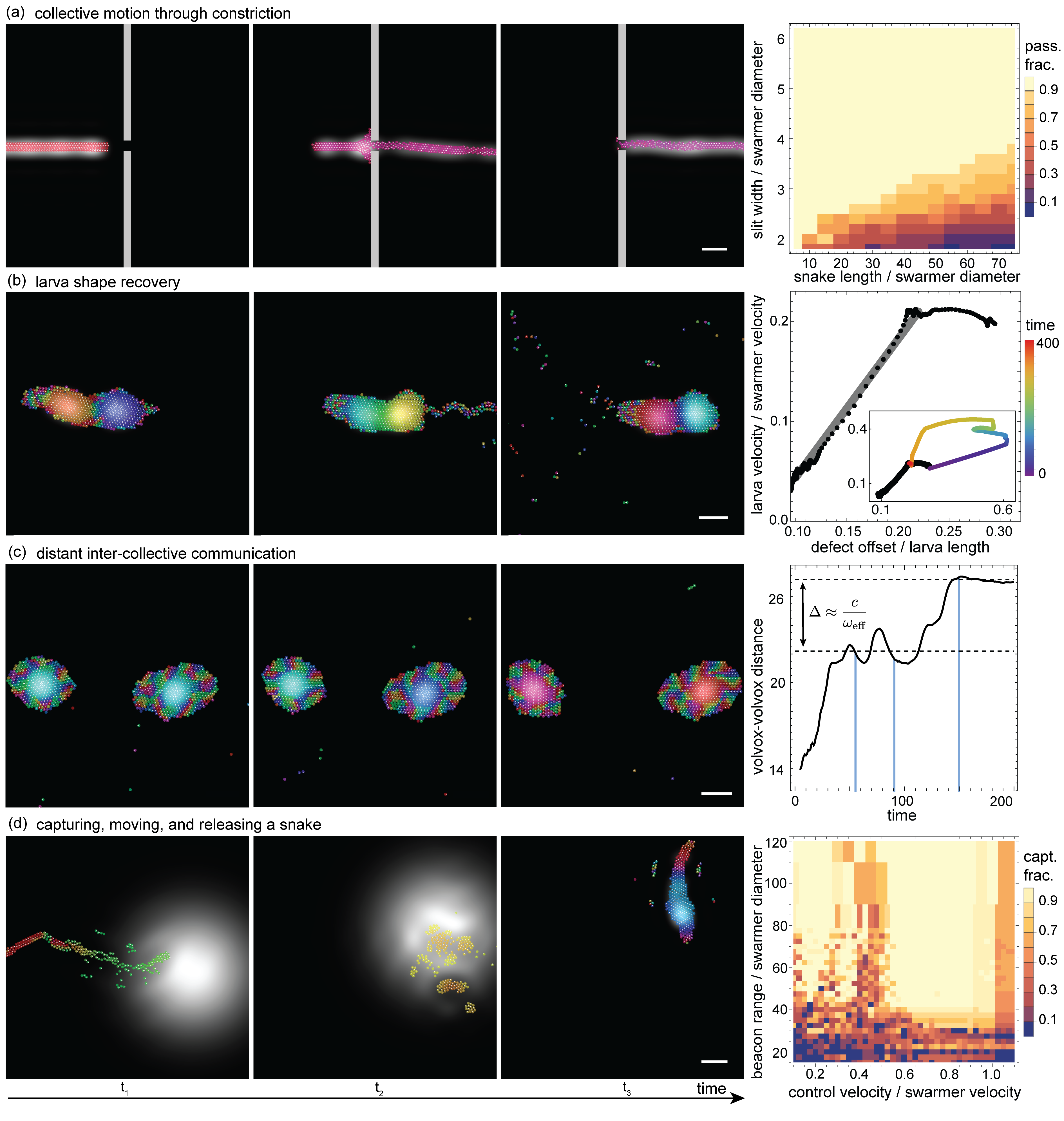}
\caption{\label{fig:function}\textbf{Illustrations of emergent functionalities.} (a) A stable snake propagates through a narrow constriction (gray). The rightmost panel shows the dependence of the passing fraction of agents on slit width and snake length. (b) A larva regrows its cut-off head. Larvae exhibit a saturating linear dependence of collective velocity on enclosed polar defect position (right panel). Upon recovery, the larva eventually returns to the projected dependence of the collective velocity of larvae states (colored trace in inset). (c) Distant acoustic communication. Two volvoxes interact via emitted acoustic waves and assume stable inter-cluster distance, an integer multiple of standing acoustic wave length $\Delta$. Blue lines indicate the times of the snapshots. (d) Capture, transport, and release of a snake by the acoustic beam (snake-in-the-egg). The applied control protocol is stable for sufficiently large beacon sizes up to control velocities equal to the individual agent velocity. The right panel indicates the fraction of successfully captured agents. Agent color code shows oscillatory phase, see Fig.~\ref{fig:pd}, and background shows normalized acoustic field amplitudes $\lvert u\rvert^2$ in grayscale. White bars indicate $5$ length units. Parameters are given in  Table \ref{tab:partPars}. See Supplementary Video 11 for temporal evolution.}
\end{figure*}

Next, we explore two other types of targeted behavior examining how systems can: (i) navigate through narrow constrictions and subsequently restore their original shape; (ii) regenerate both their shape and functionality after experiencing significant distortion.

Figure \ref{fig:function}(a) shows a time series $\left(t_1,t_2,t_3\right)$ of a snake-like collective state navigating through a narrow constriction in a wall (gray); see also Supplementary Video 11. 
Upon colliding with the wall, the snake’s shape becomes significantly distorted, suggesting that only a fraction of the agents might eventually pass through the constriction ($t_2$). 
However, after the majority of the agents successfully traversed the slit, the snake nearly regains its original shape and resumes its movement, pulling the temporarily clustered agents with it ($t_3$).
The rightmost panel in Fig.\@ \ref{fig:function}(a) shows the dependence of the agents' passing fraction on the snake length and the slit width. 
We observe that the passing fraction decreases significantly as slit width narrows or snake length increases.
For increasingly narrow constrictions, the snake must greatly reduce its width to fit through, and due to the finite size of the swarmers, more agents are left behind at the constriction. Similarly, longer snakes with more swarmers also leave behind a growing number of agents at the slit.
These left-behind agents then occasionally form a stable, localized cluster, preventing them from being pulled through with the snake, which further reduces the passing fraction. Despite these limitations, we find that snakes are often able to pass through constrictions much narrower than their initial diameter without leaving any constituting swarmers behind.

Moreover, we observe that the collective states shown in Fig.~\ref{fig:pd} exhibit shape memory and self-healing capabilities, enabling them to recover both their morphology and functionality even after experiencing strong perturbations.
For instance, if the pacemaking defect in a larva's head is destroyed (Fig.~\ref{fig:function}(b), $t_1$), the larva initially ejects some agents ($t_2$); see Supplementary Video 11. 
It then recovers by regrowing a body part that contains a new pacemaker and eventually re-absorbs the ejected agents ($t_3$). 

To quantitatively assess how the larva recovers its characteristic behavior upon the perturbation, we investigate the dependence of the aggregate's velocity on the position of the included polar defect (Fig.~\ref{fig:function}(b), right panel).
We find that the collective propagation of larvae is determined by the spatial offset of the polar defect with respect to the center of mass of the aggregate. 
As agents are oriented around the enclosed polar defect, asymmetries of the defect position control the collective velocity. 
Thus, measuring the collective velocity of various larva solutions in dependence of their defect offset (black dots, Fig.~\ref{fig:function}(b), right), we observe an approximately linear behavior up to a saturation value of about $20\,\%$ of the single swarmer's velocity (gray line). 
The linear dependence of the collective velocity on the defect localization gives a quantifiable measure for the larva phenotype.
Since the motion of all larvae is driven by the same underlying mechanism, we expect that the velocity to defect offset ratio for each larva will follow the trend shown by the black dotted line in Fig.~\ref{fig:function}(b). 
For the larva depicted in the time series (Fig.~\ref{fig:function}(b)), the strong perturbation causes its polar defect to be abruptly removed. 
In such cases where there is no enclosed polar defect, we define the defect position as being located at the outer boundary of the aggregate. Consequently, the defect offset is taken as the maximum distance from the boundary to the center of mass of the solution.
We observe that upon the strong perturbation, the aggregate leaves the characteristic linear dependence of larva velocity on defect position (trajectory, purple, inset Fig.~\ref{fig:function}(b)). 
It has lost its characteristic larva behavior and has transitioned to another phenotype with large defect offset up to $60\%$ of the larva's total length and velocities up to over $40\%$ of the individual agents. In response, the solution quickly regrows its head by developing a new enclosed polar defect. By doing so it regains the larva phenotype and returns to the characteristic velocity behavior of larvae (trajectory, red). 

In summary, we have demonstrated that, for specific swarmer characteristics and system parameters, the emergent collective states are resilient to perturbations, exhibiting a form of phenotype memory by recovering after external disruptions. 
This suggests that acoustic interactions could play a role in supporting stable, emergent functionality that remains robust to swarmer replacement while adapting to environmental changes, balancing both resilience and flexibility in collective behavior.

\subsection{Communication}

In the previous examples, we observed that the different types of collective states exhibit distinct acoustic signatures. 
As a consequence, each aggregate emits unique waves into its surroundings, enabling the potential for inter-collective communication.
An elementary example of this communication is the distance regulation of two volvox solutions (Fig.\@ \ref{fig:function}(c), Supplementary Video 11). 
Each volvox emits acoustic waves into its surrounding with individual signature similar to Fig.~\ref{fig:pd}(f). 
The interference of these emissions creates a standing wave field between the two aggregates, controlling their mutual distance.
As shown in the rightmost panel in Fig.~\ref{fig:function}(c), the inter-volvox distance oscillates about metastable distances that depend on the wavelength of the emitted acoustic signals.
Given the mutual acoustic input, the solutions have an effective average frequency $\omega_\text{eff}$, such that the standing wavelength can be approximated as
\begin{align}
    \Delta
    \approx
    \frac{c}{\omega_\text{eff}}\approx 5
    \, .
\end{align}
Thus, the dynamics of the inter-volvox distance depend on the length scale of the acoustic waves of the two aggregates, enabling them to measure their distance by means of the emitted wave strength and signal frequencies.

Beyond this very basic form of interaction studied here, acoustic communication facilitates far more complex interactions, ranging from the recognition of specific signals with orchestrated responses to the development of intricate languages. The acoustic coupling enables the exchange of information between individuals and collectives. As such it can mediate synchronization between communities depending on inherent properties and shared or distinct external stimuli.

\subsection{Position control}

Given the system's use of acoustic communication between agents, it is also responsive to external acoustic stimuli. 
As we have seen for the response to reflected signals (Sec.~\ref{sec:function_sensing}), additional inputs can give rise to significant changes in the observed behavior of the collectives. 
This external control over emergent structures has potential applications, such as manipulating the positioning, orientation, or even the exhibited phenotypes of the agents. 
In this section, we will explore how external acoustic signals can be used to control the spatial positioning of the agents.
 
Specifically, we study a scenario where we superimpose a bell-shaped acoustic signal with frequency $\omega_c$ at position $\bm{r}_c(t)$ onto the acoustic field generated by the agents: 
\begin{eqnarray}
    u_c(\bm{r},t) = A_ce^{-(\bm{r}-\bm{r}_c(t))^2/(2\sigma_c^2)}e^{i\omega_c t}
    \, ,
\end{eqnarray}
that is, we add this field as a source term to the dynamics of the acoustic field, Eq.\@ (\ref{we}).
To be specific, we select a control frequency of ${\omega_c = 2\omega_0}$, which is twice the frequency of the free agents. 
This value is chosen because it closely matches the frequencies observed in snakes, which are typically elevated by up to an order of magnitude compared to those of free agents.
However, it is important to note that this particular choice is not essential, and position control can still be effective with other frequency choices.

For capturing agents and controlling their position, the control signal, with amplitude $A_c$ and range $\sigma_c$ is moved along the protocol position 
\begin{align}
    \bm{r}_c(t)=\left\{\begin{array}{ll}
        \left(\sfrac{2}{3}\,L,\sfrac{1}{4}\,L\right)^T  & ,t<800\,,\\
        \left(\sfrac{2}{3}\,L,\sfrac{1}{4}\,L+v_ct\right)^T \phantom{sp} & ,800\leq t\leq \frac{L}{2v_c}\,,\\
        \left(\sfrac{2}{3}L,\sfrac{3}{4}L\right)^T & ,t>\frac{L}{2v_c}\,.
    \end{array}\right.
\end{align}
Using the focused acoustic beam, we aim to capture snakes and then transport them to a desired position to subsequently release them (Fig.\@ \ref{fig:function}(d), Supplementary Video 11). 
Capturing the snake at ${\bm{r}_c=(\sfrac{2}{3}\,L,\sfrac{1}{4}\,L)^T}$, we shift the target position by half system length ${L = 100}$ with velocity $v_c$.
For the snake solution in Fig.\@ \ref{fig:function}(d), the applied control first leads to the disintegration of the snake and the capture of all its agents inside the beam (\textit{snake-in-the-egg}, $t_2$). 
Moving the beam toward the target position can successfully transport the contained agents to the desired location. 
Once the beam is switched off, the agents escape the temporary confinement and readily reassemble into a snake ($t_3$). 

Even though this generic, open-loop control scheme does not require detailed knowledge of the controlled collective, it is successful over a broad range of control parameters. 
Indeed, it is effective as long as the beacon range is large enough to capture all the constituting agents and the protocol velocity does not exceed the free agent's velocity; see the rightmost panel in Fig.\@ \ref{fig:function}(d). 
However, we also observe the failure of the applied scheme, predominantly when enforcing a stronger mismatch between the control velocity and the agents' free velocity. 
This failure occurs because the agent's orientation rather than positioning is directly impacted by the control. 
For the results shown in Fig.\@ \ref{fig:function}(d), the selected control amplitudes $A_c\sim 100$ are in the same order as the acoustic field amplitudes induced by the snake.
Larger amplitude signals or more elaborate closed-loop control schemes could be applied to implement a more reliable position control. 
As the applied control initially causes the snake to transform into a blob, this generic control scheme is also effective for blob or volvox solutions (not shown here), and we anticipate potential applicability beyond the example studied in this work.

\section{Discussion and Outlook}
\label{sec:discussion}

Unlike systems at thermal equilibrium, active matter has the unique ability to form complex structures. 
This capacity arises from the constant input of energy at the level of individual agents, allowing the system to self-organize into dynamic, adaptable patterns. 
Myosin motors have been shown to generate force through the conversion of chemical energy, enabling actin filaments in actin motility assays to form self-organized, coexisting polar density waves and nematic lanes \cite{Schaller2010Polar,Huber2018Emergence}. 
Colloidal Janus particles represent another example of self-organizing active matter, as they harvest chemical energy to exhibit self-propelled motion \cite{Gomez-Solano2016Dynamics}, chemotaxis \cite{Popescu2018Chemotaxis}, and dynamically aggregate into clusters \cite{Buttinoni2013Dynamical,Niu2017Self}. 
Similarly, mixtures of actin and microtubules can self-organize into polar aster-like defects \cite{Berezney2022Extensile} and active foam structures \cite{Lemma2022Active,DeLuca2024Supramolecular}. 
These studies, along with many others, highlight the remarkable self-organization capabilities inherent to active matter systems \cite{Needleman2017Active,Bowick2022Symmetry}. 
Continuously driven out of equilibrium, these systems transform energy to enable the formation of collective structures and self-organized patterns. The building blocks of these structures are encoded in the microscopic interactions between their constituent agents.
However, to go beyond self-organization into complex structures and achieve even higher-order assemblies that can perform specific tasks or respond to environmental changes, an additional mechanism for coordination or interaction among agents is needed.
In essence, to achieve functionality, an efficient means of communication is necessary that supports both collective decision-making among agents and interaction with the environment.

Several experiments have explored the potential of chemical signaling for the development of higher-order organization into functional structures: 
It has been demonstrated that self-avoidance in self-propelled Janus particles arises from information stored within a surrounding chemical field \cite{Hokmabad2022Chemotactic}. 
Active droploids, which are aggregates of colloidal particles, exhibit self-organized polarity and coherent self-propulsion \cite{Grauer2021Active}. 
Additionally, there is a growing interest in the possibility of extracting work from active matter systems. 
For instance, when agents collectively assemble, asymmetries in their orientation and positioning can generate net rotational motion of the emergent structures \cite{Aubret2021Metamachines}. 
Passive gear-like objects, designed with chiral shapes, can act as nuclei for this aggregation and thereby promote the breaking of rotational symmetry and inducing effective rotation as a collective phenomenon \cite{sokolov2010swimming,Maggi2016Self}. 
Similarly, collective activity can be harnessed for driving active droplets \cite{Singh2020Interface}, and cargo transport \cite{Demirors2018Active}.
While such systems with chemical signaling in active matter have been well studied in recent years \cite{Theurkauff2012Dynamic,Jin2017Chemotaxis,Stark2018Artificial,Liebchen2018Synthetic,ziepke2022multi}, they rely on the diffusion of chemical species, resulting in relatively slow information exchange. 
In contrast, wave-type signaling, such as via acoustic or electromagnetic waves, offers faster communication and can readily be implemented in various synthetic or micro-robotic systems. Despite its potential advantages, this form of signaling has remained unexplored, and its role in driving emergent functional behaviors is largely unknown.

In this work, we proposed and analyzed a first theoretical model of acoustically coupled active agents, termed swarmers.
These polar agents, which self-propel in a two-dimensional habitat, are equipped with internal oscillators that continuously emit acoustic waves into their surroundings. 
Propagating through the three-dimensional space, these waves enable long-distance communication among individual agents and facilitate non-local interactions between the emergent collectives of swarmers they form.
We have coupled this long-distance information transfer to the motility of the agents by assuming that the swarmers align towards regions with higher signal amplitudes, typically where agents’ oscillations are highly synchronized.
We discuss the main results and their broader implications in the following sections.

\subsection{Emergent collective states: behavioral and acoustic phenotypes} 
Numerical simulations of this agent-based model reveal a rich diversity of emergent collective states driven by the intricate interplay of self-propulsion, mechanical alignment, and acoustic signaling. By emitting acoustic waves, the self-propelled agents synchronize their oscillations. In turn, the synchronized emission amplifies local acoustic amplitudes, creating regions with increased sound intensity. These regions then act as self-organized aggregation centers, facilitating the formation of collective structures.

We observe the formation of localized \textit{blobs}, in which swarmers cluster around a highly synchronized central region containing a polar defect. 
Additionally, we identify \textit{larva} solutions, which move slowly and exhibit an asymmetrical arrangement of agents around a polar defect. Yet another collective structure we observe are fast-moving \textit{snakes}, which consists of a cohesive group of swarmers that are well-aligned and move coherently in a common direction. 
These snake solutions bear a strong resemblance to structures in polar active matter with vision cone alignment \cite{negi2024collective}. 
There, a non-reciprocal interaction between agents is globally imposed by varying vision cones that control the mutual alignment interaction \cite{barberis2016large,negi2024collective}. 
In contrast, our model features emergent symmetry breaking, with a self-organizing pacemaker acting as a phase leader. 
It guides the propagation of phase waves through the structure and thereby controls directed information transport. Finally, acoustic active matter exhibits \textit{volvox} structures with the synchronized central region and an outer layer of incoherently oscillating swarmers and ring-like \textit{ouroboros} structures. 
The volvox solutions resemble chimera states in networks of non-locally coupled oscillators characterized by the coexistence of synchronized and incoherently oscillating agents \cite{kuramoto2002coexistence,abrams2004chimera}. 
In contrast to static networks of oscillators, volvoxes are spatially organized chimera structures with a dynamic exchange of interaction partners due to the self-propulsion of agents. 
Their synchronized center and the incoherently oscillating outer layers reflect two opposed regimes of acoustic signaling. 
These could be further linked to different behavioral modes to gain higher order functionality. 
As the incoherence in the agents identifies the positioning within the outer layer of the aggregate, agents could develop properties important for a specific physical interaction. 
For instance, they could develop into a static shell, further protecting the aggregate.

Taken together, the acoustic active matter is characterized by a rich variety of phenotypes, each exhibiting distinct morphology and mesoscopic behavior.
The occurrence of the different phenotypes is closely influenced by the chosen values of microscopic agent parameters. 
Our analysis shows that slow, highly susceptible swarmers form localized blobs, while fast, persistent agents generate snake-like assemblies. The other solutions arise in intermediate regimes. This suggests that manipulating these microscopic parameters allows for a control of the emergent phenotypes. In this study, we focused on emergent functionality based on basic alignment with acoustic amplitude gradients. However, designing more complex functional structures will increasingly require incorporating higher-order interactions between microscopic agent parameters and the information derived from the signaling field.

The emergent structures not only show distinct behavioral phenotypes but also produce specific acoustic emission characteristics. 
Each collective state generates different frequency distributions and emission amplitudes, making structures distinguishable based on their acoustic signatures. Through the emission of these distinctive signals, the collectives can communicate with one another, allowing them to identify the type and potentially also the current state of the surrounding solutions. 
As a result, the collectives can develop varied behavioral responses to the different acoustic interactions they encounter.

\subsection{Field theory of acoustic active matter}

Beyond exploring emergent structures, we also investigated how the aggregation process continues on larger time- and length scales. To access these, we proposed complementary continuous field equations as a coarse-grained version of the agent-based model. Similar to known hydrodynamic field equations for polar active matter \cite{Toner1995Long,aranson2005pattern,Bertin2006Boltzmann,Baskaran2008Hydrodynamics,Julicher2018Hydrodynamic}, our model includes polar alignment interactions of swarmers by a density-dependent isotropic-to-polar order transition and the alignment with gradients in the acoustic signaling amplitude. This field-based approach successfully reproduces the blobs and snake solutions as two central collective states of the system. Thereby, the model captures the essentials of the structures’ phenotypical and acoustic signatures. Over longer time scales, we have observed rapid coarsening of aggregates, much faster than classical Ostwald-ripening or cluster aggregation in motility-induced phase separation. Ultimately, clustering stabilizes at a length scale that is influenced by the acoustic wavelength. We find that different clusters interact with each other via the acoustic field and extract distance and positional information from it. 
This highlights once again the significant role of wave-like coupling in cluster aggregation and long-distance information transfer. 
The hydrodynamic field theory serves as an effective tool to study acoustic coupling in very large systems and may be valuable for future studies of systems with large agent numbers. 
Our study demonstrates that information can be propagated efficiently over large distances. While we focused on system sizes where wave absorption could be neglected, absorption characteristics can vary across different media. Future studies with specific settings may need to consider such additional factors like wave absorption and boundary reflections.

\subsection{Adaptive responses and interactions with the environment}

For the emergence of collective functional structures, adaptability to environmental changes is crucial as it enables the collectives to locate target positions and to organize and monitor their function. 
It is widely recognized as critical for the functioning of biological systems \cite{Garnier2007Biological}, spanning from complex animals such as social insects~\cite{Detrain2008Collective} to microscopic organisms like bacteria~\cite{Ben-Jacob2006Self}. However, despite its importance,  few studies have investigated the capabilities of synthetic active matter systems to respond to environmental changes. For instance, active colloids can form swirl-like structures and adapt their self-propulsion in response to external changes \cite{Bauerle2020Formation}. More broadly, colloidal and de-mixed droplet systems can measure and respond to changes in their chemical or optical surroundings, and adapt their self-propulsion accordingly \cite{Aubret2021Metamachines,Hokmabad2022Chemotactic,Jacucci2024Patchy}. These initial implementations of environmental sensing in synthetic systems already demonstrate its potential impact on the systems' collective behavior. 
Unlike these previous studies that rely on individual agents to sense environmental cues, our results demonstrate a collective form of environmental sensing in acoustic active matter that emerges as a cooperative function through acoustic synchronization within the collectives.
In our model, the environmental sensing is achieved through the acoustic field. Agents emit synchronized waves and detect reflections from objects in their surroundings. 
Thereby, the swarmers gain a cooperative increase in the strength of environmental coupling by collectively emitting stronger signals. 
We have examined how blob and larva structures respond to an approaching reflective object above their habitat. The reflected signals trigger phenotype changes in the solutions and lead to cluster localization or dispersal of agents. This demonstrates that collectives can sense and react to external stimuli using the acoustic field. 
Similarly, inter-cluster communication is facilitated by acoustic signals. We observed that two volvox-like aggregates could maintain a stable distance by sensing and responding to each other's acoustic emissions, essentially measuring their separation through these interactions.

Our study also reveals that different collective functions require distinct behaviors of the individual swarmers. 
In a system of identical units, the swarmers must adapt and differentiate their behaviors in response to the acoustic input they receive from their surrounding. 
As this behavioral differentiation is self-organized and agents are functionally identical, the system can compensate for agent failures and dynamically adapt to imposed perturbations. 
As a result, the collective states acquire a high level of robustness.
For example, snake structures can collectively navigate through constrictions narrower than their original diameter, and larva structures can recover their polar defect, demonstrating their resilience.

 Finally, we presented that acoustic waves can be used to externally control the system. For instance, snake structures can be captured and relocated via acoustic signals, resuming their behavior once released at the desired position. Altogether, this enables external supervision and control of the active matter system through the measurement and application of acoustic signals.

In conclusion, we have shown that wave coupling between self-propelled active agents yields various distinct functional structures with emergent capabilities. The emission and detection of acoustic signals by the swarmers enable a fast information exchange over large distances. Through the acoustic field, aggregates communicate characteristic acoustic signals and gain information about their environment via a collective sonar-like mechanism. 

\subsection{Towards a cybernetics of active matter}

We believe that the present study takes an important step
towards a new form of active matter that is able to organize into collective states that can be regarded as phenotypes, which exhibit higher-level features (functions) that allow them to respond in an adaptive way to changes in the environment.
In the proposed framework of acoustic active matter, the system acquires emergent functionality without external supervision, relying solely on microscopic interaction rules at the level of individual swarmers. 
Similar to neural networks, collective functionality arises from the interaction among units, with each performing only simple computational steps. This approach keeps individual agents simple and their computational energy consumption minimal, as each swarmer processes only a small portion of the information available to the entire cluster.

Our minimal model for wave interaction already demonstrates the rich collective behavior achievable through acoustic coupling. This model offers insights into the fundamental principles of wave-type interactions and the collective organization of oscillatory self-propelled units, as well as the emergence of collective sensing. It serves as a foundation for exploring more complex interactions and microscopic behaviors, guiding the development of functional active matter systems toward more specific and advanced applications.

While our model highlights the rich collective behavior enabled by acoustic interactions between oscillatory units, practical implementations may face several challenges that require careful adaptation. Robotic agents, for instance, must incorporate directional acoustic sensing, either by employing at least two microphones or by temporally sampling local acoustic field amplitudes. Similarly, electronic circuits may exhibit oscillatory behaviors that deviate from the reduced Stuart-Landau form, necessitating either alternative designs or emulation via microprocessors. Additional physical effects, such as signal absorption or reflection by other agents, may further complicate real-world applications. Despite these challenges, our study demonstrates the potential of acoustic signaling as a robust mechanism for collective self-organization, long-range information transfer, and adaptive responses to environmental changes.

The potential applications for unsupervised functional active matter systems are diverse. Once this form of communication is integrated into synthetic systems, it could enable tasks in environments that are otherwise inaccessible or hazardous. The robustness and adaptability of these structures suggest significant potential for real-world applications in environments where external supervision is impractical.
For example, following the presented principles, ensembles of acoustically communicating agents may develop a more evolved cooperative sonar, where phase differences in reflected signals can be evaluated by the collective, yielding insights into the nature of the reflective objects and triggering appropriate behavioral responses.
The wave-like coupling has been motivated by acoustic waves and could be used by naval drones or robots in a mechanically coupled medium \cite{baconnier2023discontinuous,xu2023autonomous,pratissoli2023coherent}. One can expect similar behavior from electromagnetically communicating agents.

From a broader perspective, future studies should extend beyond the microscale energy conversion---that defines the field of active matter---to explore the ability of agents to perceive and respond to their environment.
Through communication, these agents can then form collectives that exhibit cooperative behavior, make collective decisions, and actively reshape their surroundings. 
Investigating such functional synthetic active matter lays the groundwork for cybernetics of active systems, which focuses on designing and controlling synthetic systems to achieve specific objectives.

\begin{acknowledgments}
We thank Jo\"{e}l Schaer for insightful discussions and comments.
ISA and EF thank the support of the John Templeton Foundation, award 63296. 
EF acknowledges financial support from the German Research Foundation (DFG) through the Excellence Cluster ORIGINS under Germany's Excellence Strategy (EXC 2094 -- 390783311). EF was supported by the European Union (ERC, CellGeom, project number 101097810) and the Chan--Zuckerberg Initiative (CZI).
\end{acknowledgments}

\clearpage\newpage

\appendix
\counterwithin*{figure}{part}

\stepcounter{part}

\renewcommand{\thefigure}{A\arabic{figure}}

\section{Quasistatic solution of the wave equation}\label{sec_app:static}
The system of acoustic active matter introduces a wave-like coupling between self-propelled oscillators. In the model, the swarmers, confined to their two-dimensional habitat, emit waves into the surrounding three-dimensional environment. Typically, the acoustic wavelengths are much larger than the size of the individual agents, and the speed of sound significantly exceeds the self-propulsion velocities of the swarmers. In this section, we explain how we take advantage of these scale differences to incorporate acoustic coupling by employing a quasi-static solution to the three-dimensional wave equation with the agents as acoustic sources.
The dynamics of the sound field $u(\bm{r},t)$ in three spatial dimensions, $\bm{r}=\left(x,y,z\right)^T\in\mathbb{R}^3$, is modeled using the wave equation
\begin{align}\label{eq_sm:wave_eq}
    \frac{1}{c^2}\partial_t^2 u(\bm{r},t)&=\nabla^2 u+\sum_j w(\bm{r}-\bm{r}_j,t)a_j(t)\delta(z)\,,
\end{align}
with the speed of sound $c$ and active agents as sources confined to the two-dimensional plane at $z=0$. As detailed in the main text, the agents' source contribution $a_j(t)$ is weighted by a Gaussian kernel $w(\bm{r},t)$, representing the agent's size. Assuming fast sound dynamics as wave propagation is fast compared to the agent velocity, i.e., $c\ll v_0$, we consider the quasi-stationary case, 
\begin{align}
    c^2\nabla^2 u(\bm{r})+g\delta(z) = 0\,.
\end{align}
with general source contributions $g\delta(z)$. It can be solved using a separation ansatz for the in-plane solution in the two-dimensional habitat and the perpendicular out-of-plane direction, $u(\bm{r})=u_{x,y}(x,y)Z(z)$. Assuming periodic boundary conditions within the habitat, we apply a Fourier transform in $(x,y)$-plane with two-dimensional wave vector $\bm{k}=(k_x,k_y)^T$. Then, outside the habitat plane ($z\neq 0$) in the absence of additional sources, the stationary wave equation reads
\begin{align}
0&=c^2\left(-k_x^2-k_y^2+\partial_z^2\right)\tilde{u}_{\bm{k}}Z(z)\,,
\end{align}
with Fourier transform $\tilde{u}_{\bm{k}}\equiv\mathcal{F}\left[u_{x,y}(x,y)\right](\bm{k})$.
Next, we aim to find the corresponding out-of-plane bulk solution $Z(z)$.
For non-vanishing in-plane wave vectors, $k=\lvert\bm{k}\rvert=\left(k_x^2+k_y^2\right)^{1/2}\neq 0$, the equation reads
\begin{align}
   \partial_z^2Z(z)&=\left(k_x^2+k_y^2\right)Z(z)\,,\\
   \intertext{and one can find the solution}
   Z(z)&=e^{-k\lvert z\rvert}\,,
   \end{align}
   which fulfills the boundary conditions of vanishing contributions at $z\rightarrow\pm\infty$.
    Waves emitted from the two-dimensional habitat exponentially decay in the bulk depending on their wavelength $1/k$. Accordingly, for the homogeneous emission from the bulk, $k=0$, the equation reduces to
   \begin{align}
   \partial_z^2Z(z)&=0\,,\\
   \intertext{with the solution}
   Z_0(z)&=\alpha+\beta\lvert z\rvert\,.
\end{align}
As we neglect absorption in the medium, which is valid for sufficiently small systems, the decay of waves away from the habitat is only caused by interference effects between the waves emitted at different positions within the two-dimensional plane. The waves will overlay and eventually average out at some height above the habitat. Homogeneous oscillations, on the other hand, will never average out as they have a net contribution over the whole domain.
Next, we consider the sources at $z=0$, such that the full solution to the wave equation has to fulfill
\begin{align}
    c^2\left(\partial_z^2-k^2\right)\tilde{u}_{\bm{k}}Z(z)+\tilde{g}_{\bm{k}}\delta(z)=0\,.
\end{align}
We integrate this equation over a small interval around the habitat $z\in\left(-\epsilon,\epsilon\right)$ and subsequently consider the limit $\epsilon\rightarrow 0$. As, the integral
\begin{align}
\lim_{\epsilon\rightarrow 0}\int_{-\epsilon}^{\epsilon}Z(z)k^2\mathrm{d}z =0
\end{align}
vanishes in the limit, we remain with the terms
\begin{align}\label{eq:app_limit}
    \lim_{\epsilon\rightarrow 0}c^2\partial_z\tilde{u}_{\bm{k}}Z(z)\Big\rvert_{-\epsilon}^{\epsilon}+\tilde{g}_{\bm{k}}=0\,.
\end{align}
 Evaluating the limit, we obtain for $k\neq0$,
\begin{align}
   -2c^2\tilde{u}_{\bm{k}}kZ(z)+\tilde{g}_{\bm{k}}&=0\,,
   \end{align}
   which is solved by
   \begin{align}
   \tilde{u}_{\bm{k}}Z(z)&=\frac{\tilde{g}_{\bm{k}}}{2kc^2}\,.
   \end{align}
   For vanishing wave vectors, $k=0$, evaluation of the limit in equation, Eq.~(\ref{eq:app_limit}), yields
   \begin{align}
2c^2\tilde{u}_0\beta+\tilde{g}_0&=0\,,
\end{align}
with the solution
\begin{align}
    \tilde{u}_0Z_0(z)&=\tilde{u}_0\left(\alpha-\frac{\tilde{g}_0\lvert z\rvert}{2c^2\tilde{u}_0}\right)\,.
\end{align}
Finally, as argued before, we neglect any homogeneous contributions to the acoustic field, $\tilde u_0=0$ as these would represent global offsets and therefore just shift the baseline of the acoustic field. Consequently, for a given Fourier representation of the sources, $\tilde{g}_{\bm{k}}$, the acoustic field at the habitat, $z=0$, is given by
\begin{align}\label{eq_app:a_solution}
    \tilde{u}_{\bm{k}}=\frac{\tilde{g}_{\bm{k}}}{2c^2\sqrt{k_x^2+k_y^2}}\,.
\end{align}
Beyond this quasi-stationary limit, one can assume that the temporal evolution of the acoustic field is governed by a predominant frequency $\omega_u$ close to the intrinsic frequency of agents, $\omega_u\approx\omega$, with acoustic field $u(t)\sim e^{i\omega_u t}$. Then, an extended version of the solution, Eq.~\eqref{eq_app:a_solution}, is given by
\begin{align}
    \tilde{u}_{\bm{k}}=\frac{\tilde{g}_{\bm{k}}}{2c^2\sqrt{k_x^2+k_y^2-\omega_u^2/c^2}}\,.
\end{align}

Neglecting damping, we solve the three-dimensional quasi-static wave equation for acoustic emitters confined to a two-dimensional plane. The resulting acoustic field gives rise to a non-local acoustic coupling that scales as $1/k$ in Fourier space. To understand the impact of this non-local coupling on structure formation, we study a minimal version of the acoustic active matter. Namely, for homogeneous agent densities, the system simplifies to a non-locally coupled complex Ginzburg-Landau equation, see App.~\ref{app:cgle}.

\section{Non-locally coupled complex Ginzburg Landau equation}\label{app:cgle}
In the main text, we discuss how for constant densities, $\rho\equiv\rho_0$, the system of acoustic active matter, Eq.~(\ref{eq:agents}), can be described by a non-locally coupled version of the complex Ginzburg Landau equation (CGLE). Indeed, for homogeneous densities and without swarmer motility, $\rho(x,y)=\rho_0$, $v_0=0$, the system reduces to
{\small\begin{align}
    \partial_t a&=\mu\nabla_{2D} a+\left(1+i\omega\right)a-\left(1+ib\right)\lvert a\rvert^2a+\lambda u\lvert_{z=0}\label{eq:dyn_a_app}\\
    \frac{1}{c^2}\partial_t^2 u &=\nabla^2u+\rho_0 a\delta(z)\,.
    \label{app_eq:b_wave}
\end{align}
}

Assuming large wave propagation velocities $c\gg 1$, one can consider the quasi-static wave equation, namely the Poisson equation,
\begin{align}
    \nabla^2u=-\rho_0a\delta(z)\,,
\end{align}
with solutions given by Eq.~(\ref{eq_app:a_solution}), see App.~\ref{sec_app:static}. Inserting this solution into Eq.~(\ref{eq:dyn_a_app}), we consider plain wave solutions,
\begin{align}
    a(x,t)=a_0\exp\left\{i\Omega t+ikx\right\}\,.
\end{align}
Plugging this ansatz into Eq.\@ (\ref{eq:dyn_a_app}) yields a dispersion relation $\Omega(k)$,
\begin{align}
    i\Omega&=\left(1+i\omega\right)-\left(1+ib\right)a_0^2+\frac{\lambda\rho_0}{2\lvert k\rvert}-\mu k^2\,.
\end{align}
From it, one can see that the coupling of the oscillatory medium mediated by the acoustic field induces non-local interactions ($\sim 1/\lvert k\rvert$) which scale with the wavelength of the considered wave. Solving this equation, the amplitude and, respectively, the phase of the plane-wave solutions read
\begin{align}
    a_0&=1+\frac{\lambda\rho_0}{2\lvert k\rvert}-\mu k^2\\
    \Omega&=\omega-ba_0^2\,.
\end{align}
As for the classical CGLE, the frequency of the acoustic wave has a non-linear relationship with the amplitude as $\Omega\sim ba_0^2$. The non-local acoustic coupling reflects itself in an additional contribution ($\sim\lambda$) which modulates the plane wave amplitude.
Considering the group velocity within the plain wave solution, one gets
\begin{align}
    v_g=\frac{\partial\Omega}{\partial k}=2ba_0\left[\frac{\lambda\rho_0 k}{2\lvert k\rvert^3}-2\mu k\right]\,.
\end{align}
As such, the acoustic coupling induces a divergence of the group velocity for long wavelength modes, $k\rightarrow 0$. As discussed in App.~\ref{sec_app:static}, this is due to the fact that waves overlay and propagate through the three-dimensional environment. They eventually annihilate over length scales at which positive and negative contributions vanish on average. For larger wavelength, $k\rightarrow 0$, this length scale over which waves negatively interfere becomes larger and larger. Thus, interactions become increasingly strong for larger wavelengths of the acoustic signals. This non-local coupling on large length scales has a significant impact on the system's coarsening dynamics, as we will see below.
\begin{figure}[!tb]
    \centering
\includegraphics{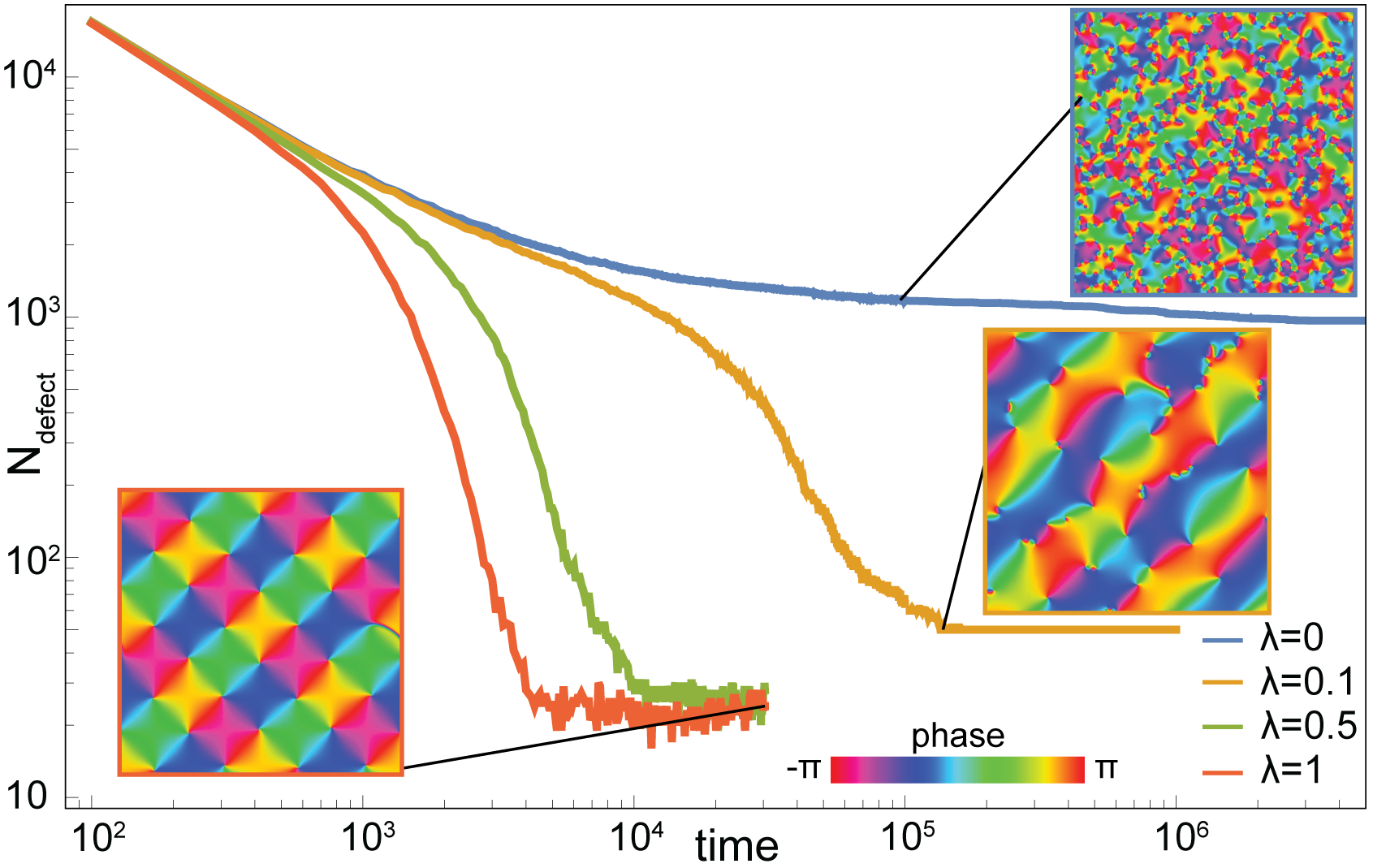}
    \caption{\textbf{Temporal evolution of number of phase defects} $N_{\text{defect}}$\textbf{.} The non-local coupling $\lambda>0$ accelerates defect coarsening significantly until a final length scale is reached. Measured numbers correspond to a box with side length $L=1000$ with periodic boundary conditions. System parameters as in Table \ref{tab:fieldPars}}
    \label{fig:defect_coarsening}
\end{figure}
We measure the number of phase defects in the non-locally coupled CGLE, Eq.~(\ref{eq:dyn_a_app}), for a constant homogeneous agent density $\rho_0= 0.6$ over time (Fig.\@ \ref{fig:defect_coarsening}). Without the long-range acoustic coupling, $\lambda=0$, we observe slow defect coarsening due to the exponentially decaying interaction between defects \cite{aranson2002world,brito2003vortex}. In contrast, enabling acoustic coupling yields a significantly faster merging of defects as regions synchronize via the acoustic field. The final number of defects, $N_{\rm defect}$,  saturates according to the CGLE screening length, $L_{\rm screening}\sim 1/b k_0$, with the wavenumber selected by the defects $k_0$. The explicit relationship can be obtained from the asymptotic linear stability of the planar waves emitted by the topological defects \cite{aranson2002world}. Then, one can find that the defect number $N_{\text{defect}} \sim L^2 /L_{\text{screening}}^2$, where $L$ represents the system size \cite{aranson2002world}. For conciseness, we will not perform the calculations explicitly here.
The study of this minimal system for acoustically non-locally coupled oscillators emphasizes the significance of the long-range interaction for the coarsening of defects on a large scale.

\section{Quantification of phenotypes}
In this section, we provide a detailed analysis of the representative phenotypical states discussed in the main text section \ref{sec:States}. In particular, we examine the agents' oscillatory phases
\begin{align}\label{app_eq:phases}
    \phi_j=\arctan\left(\frac{{\rm Im}\left(a_j\right)}{{\rm Re}\left(a_j\right)}\right)\,,
\end{align}
the direction of the agents' polar orientation $\bm{p}_j$, and the spatial distribution of frequencies of the oscillatory units. Additionally, we use the local Kuramoto order parameter \cite{KuramotoChemical}
\begin{align}\label{app_eq:Kuramoto_local}
    z_j=\frac{1}{N_j}\bigg\lvert\sum_{k,\,\lvert\bm{r}_j-\bm{r}_k\rvert<5r_\text{p}}e^{i\phi_k}\bigg\rvert\,,
\end{align}
as a measure for local synchronization, averaging oscillatory states in an environment around the particular agent $j$.
\subsection{Blob state}
\begin{figure}
    \centering
    \includegraphics[width=\linewidth]{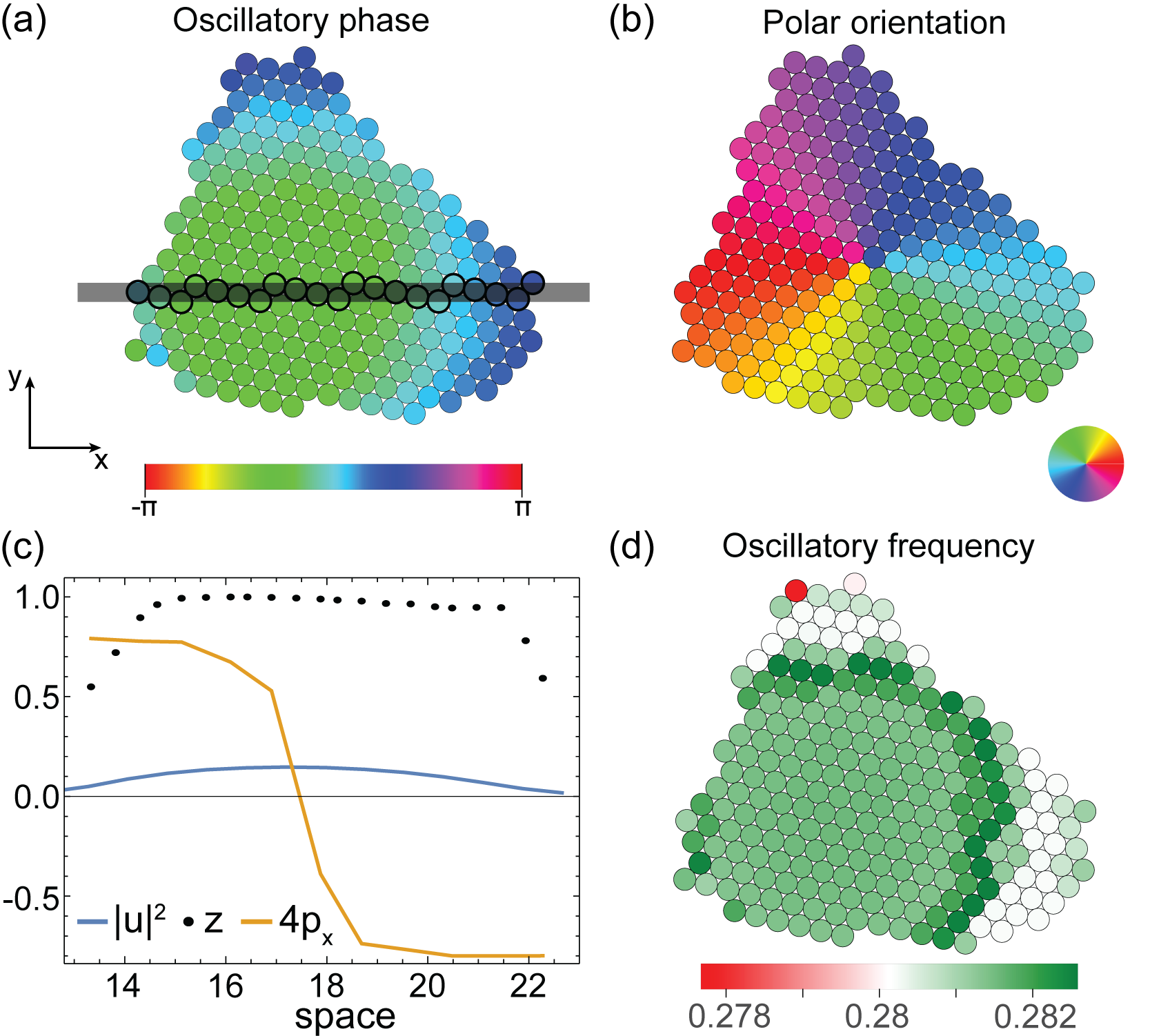}
\caption{\textbf{Detailed analysis of blob state.} (a) Agents' oscillatory phases, Eq.~\eqref{app_eq:phases}. (b) Polar orientation of agents. (c) Quantification of acoustic field amplitudes $\lvert u\rvert$, local Kuramoto order parameter $z$, Eq.~\eqref{app_eq:Kuramoto_local}, and polar orientation in $x$ direction $p_x$ for agents along the gray shaded line indicated in panel (a). (d) Oscillatory frequencies. Parameters as stated in Table \ref{tab:partPars} for Fig.~\ref{fig:pd}.}
    \label{app_fig:blob}
\end{figure}

As stated in the main text, blob solutions are characterized by concentric phase waves, Fig.~\ref{app_fig:blob}(a) and polar orientation of agents towards the center of the aggregate, Fig.~\ref{app_fig:blob}(b). Figure \ref{app_fig:blob}(c) quantifies the spatial distribution of acoustic field amplitudes $\lvert u\rvert$, the local Kuramoto order parameter $z$, Eq.~\eqref{app_eq:Kuramoto_local}, and the polar orientation in $x$-direction, $p_x$ for agents along the gray shaded line indicated in figure panel (a).
The data confirms that acoustic amplitudes are largest in the center of the blob solution, and the polar orientation exhibits a steep transition in the blob's center area, indicating the central polar defect in the structure. Moreover, the Kuramoto order parameter $z$ attains maximal values $z\lesssim 1$ almost throughout the entire aggregate, representing the high level of synchronization within blob solutions. Only at the outermost regions does synchronization weaken, $z\approx0.5$, revealing the presence of characteristic concentric phase waves. These are also visible in the spatial frequency distribution, where we observe a slight increase of frequencies in the blob's center region due to the stronger acoustic coupling and higher level of synchronization.
\subsection{Larva state}
\begin{figure}
    \centering
    \includegraphics[width=\linewidth]{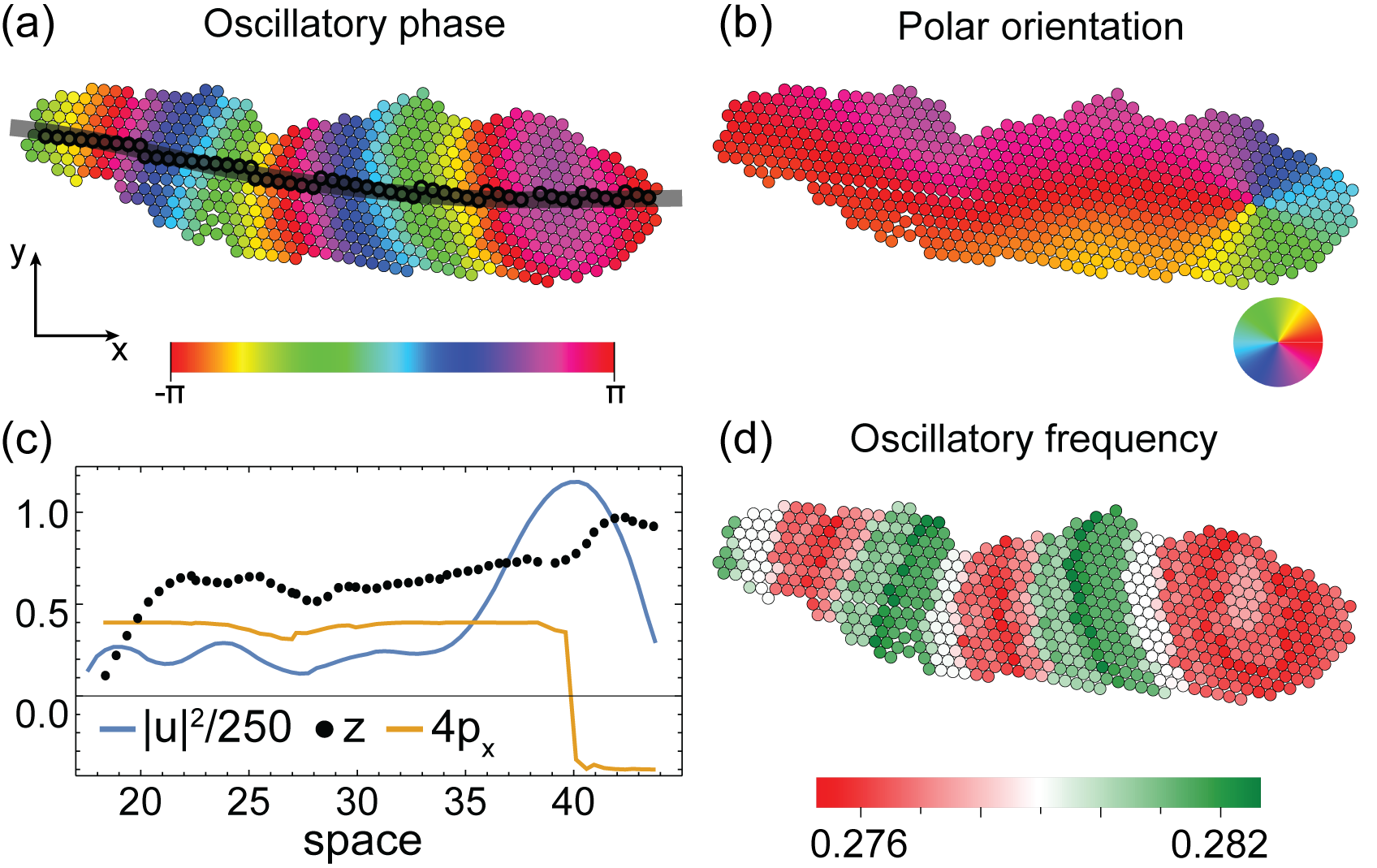}
    \caption{\textbf{Detailed analysis of larva state.} (a) Agents' oscillatory phases, Eq.~\eqref{app_eq:phases}. (b) Polar orientation of agents. (c) Quantification of acoustic field amplitudes $\lvert u\rvert$, local Kuramoto order parameter $z$, Eq.~\eqref{app_eq:Kuramoto_local}, and polar orientation in $x$ direction $p_x$ for agents along the gray shaded line indicated in panel (a). (d) Oscillatory frequencies. Parameters as stated in Table \ref{tab:partPars} for Fig.~\ref{fig:pd}.}
    \label{app_fig:larva}
\end{figure}

Larvae are characterized by phase waves propagating from head to tail, Fig.~\ref{app_fig:larva}(a), and consist of an asymmetric aggregation of agents around a polar defect in their head, Fig.~\ref{app_fig:larva}(b).
Considering the acoustic field amplitudes along the centerline, indicated as the gray-shaded region in Fig.~\ref{app_fig:larva}(a), we observe the largest field amplitudes in the head region. This is caused by a co-localized region of strongest synchronization, as can be seen from the Kuramoto order parameter $z$, Fig.~\ref{app_fig:larva}(c). Characteristic of a larva, the polar orientation of agents exhibits a steep transition in the larva's head, signifying the enclosed polar defect. The oscillatory frequencies show a slight modulation throughout the larva structure that corresponds to the spatial distribution of acoustic field amplitudes, Fig.~\ref{app_fig:larva}(d).
\subsection{Snake state}
\begin{figure}
    \centering
    \includegraphics[width=\linewidth]{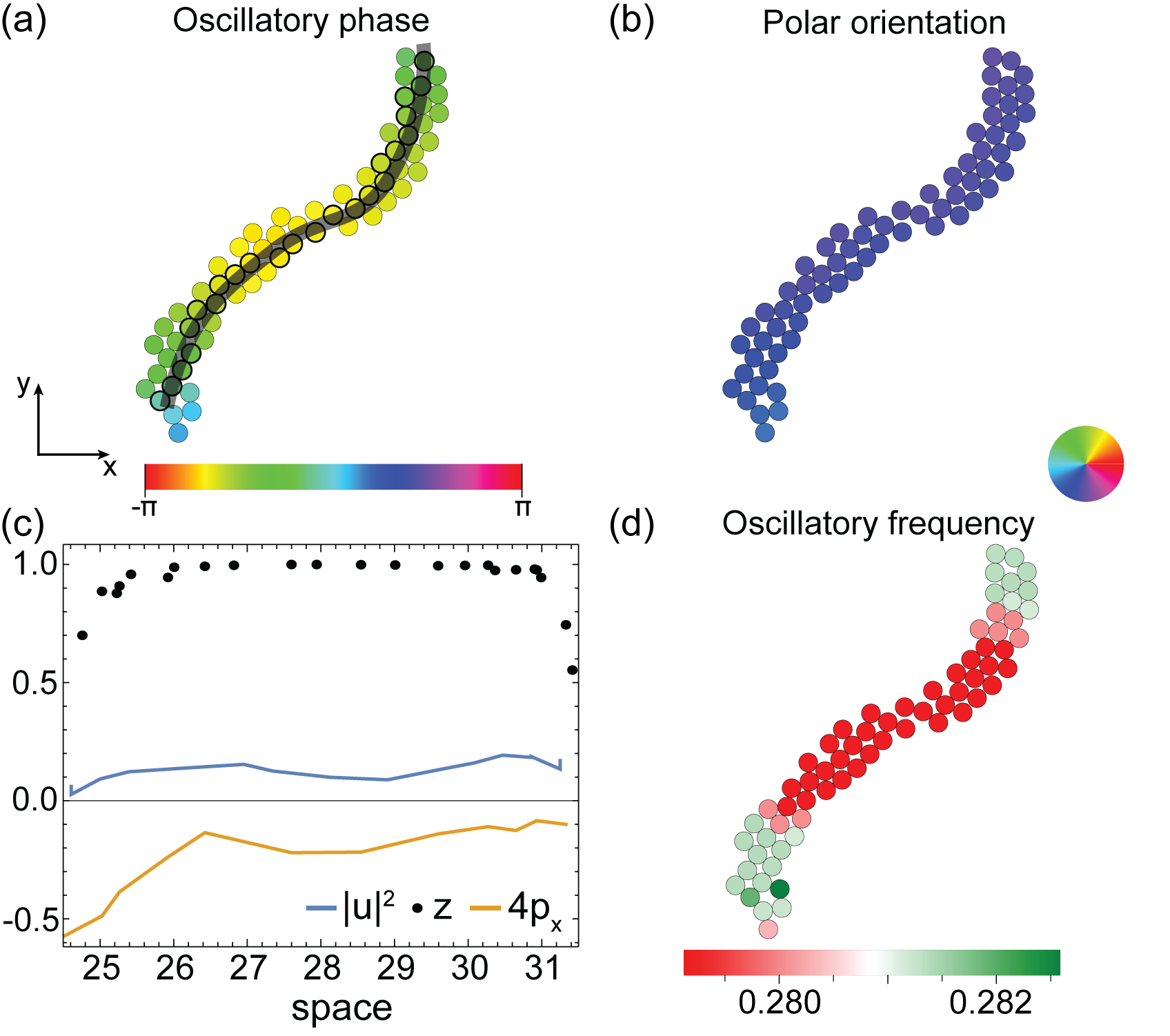}
    \caption{\textbf{Detailed analysis of snake state.} (a) Agents' oscillatory phases, Eq.~\eqref{app_eq:phases}. (b) Polar orientation of agents. (c) Quantification of acoustic field amplitudes $\lvert u\rvert$, local Kuramoto order parameter $z$, Eq.~\eqref{app_eq:Kuramoto_local}, and polar orientation in $x$ direction $p_x$ for agents along the gray shaded line indicated in panel (a). (d) Oscillatory frequencies. Parameters as stated in Table \ref{tab:partPars} for Fig.~\ref{fig:pd}.}
    \label{app_fig:snake}
\end{figure}

Snake states exhibit various forms of phase waves as exemplarily shown in Fig.~\ref{app_fig:snake}(a), and agents are approximately aligned in a polar fashion along a common direction of motion, Fig.~\ref{app_fig:snake}(b).
Given the phase wave along the snake’s centerline, the structure is characterized by relatively low and only weakly varying acoustic amplitudes, while exhibiting high synchronization values, $z \gtrsim 0.8$, and strong polar alignment, Fig.~\ref{app_fig:snake}(c). The oscillatory frequencies display minimal variation, consistent with the weak modulation of both acoustic amplitudes and oscillatory synchronization levels.
\subsection{Ouroboros state}
\begin{figure}
    \centering
    \includegraphics[width=\linewidth]{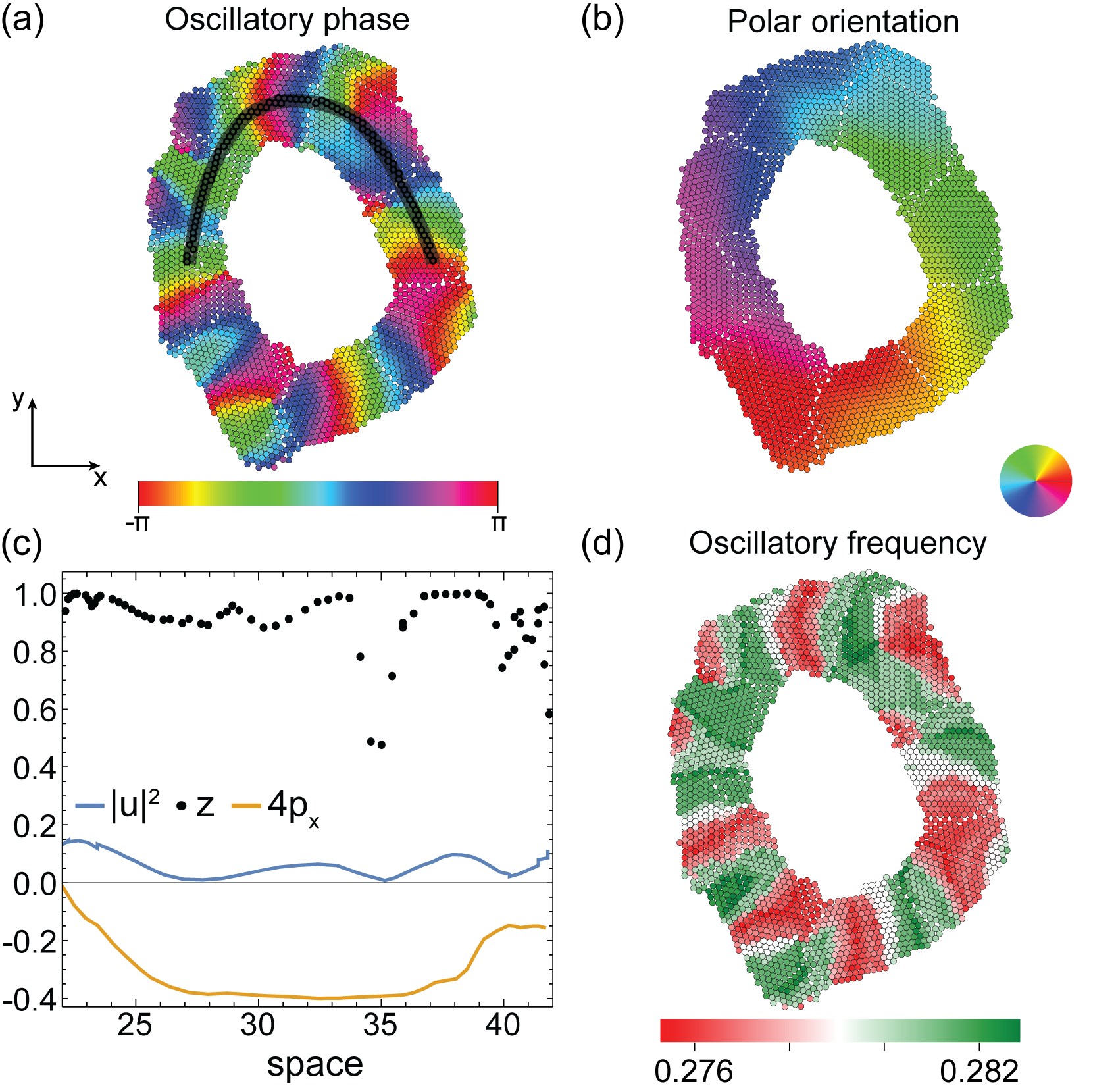}
    \caption{\textbf{Detailed analysis of ouroboros state.} (a) Agents' oscillatory phases, Eq.~\eqref{app_eq:phases}. (b) Polar orientation of agents. (c) Quantification of acoustic field amplitudes $\lvert u\rvert$, local Kuramoto order parameter $z$, Eq.~\eqref{app_eq:Kuramoto_local}, and polar orientation in $x$ direction $p_x$ for agents along the gray shaded line indicated in panel (a). (d) Oscillatory frequencies. Parameters as stated in Table \ref{tab:partPars} for Fig.~\ref{fig:pd}.}
    \label{app_fig:ouroboros}
\end{figure}

Ouroboros states are characterized by closed, ring-like structures exhibiting various phase-wave patterns, including phase defects, Fig.~\ref{app_fig:ouroboros}(a). The agents align along the ring, generating a net rotational motion of the structure, Fig.~\ref{app_fig:ouroboros}(b). However, the circular movement of agents is partially obstructed by the agents ahead, resembling the behavior of larvae. However, in the case of ouroboroi, the structures form a closed loop.
Examining the acoustic field amplitudes along the gray-shaded agents (see Fig.~\ref{app_fig:ouroboros}(a)), the ouroboros state exhibits significant modulations, Fig.~\ref{app_fig:ouroboros}(c). This arises from the presence of oscillatory phase defects, where acoustic amplitudes locally cancel out. A clear correspondence is observed between these phase defects and regions of significantly decreased local Kuramoto order parameters, $z\lesssim 0.6$, as well as near-vanishing acoustic field amplitudes. Due to these variations in amplitude and synchronization, the agents’ frequencies also fluctuate throughout the structure, Fig.~\ref{app_fig:ouroboros}(d).

\subsection{Volvox state}
\begin{figure}
    \centering
    \includegraphics[width=\linewidth]{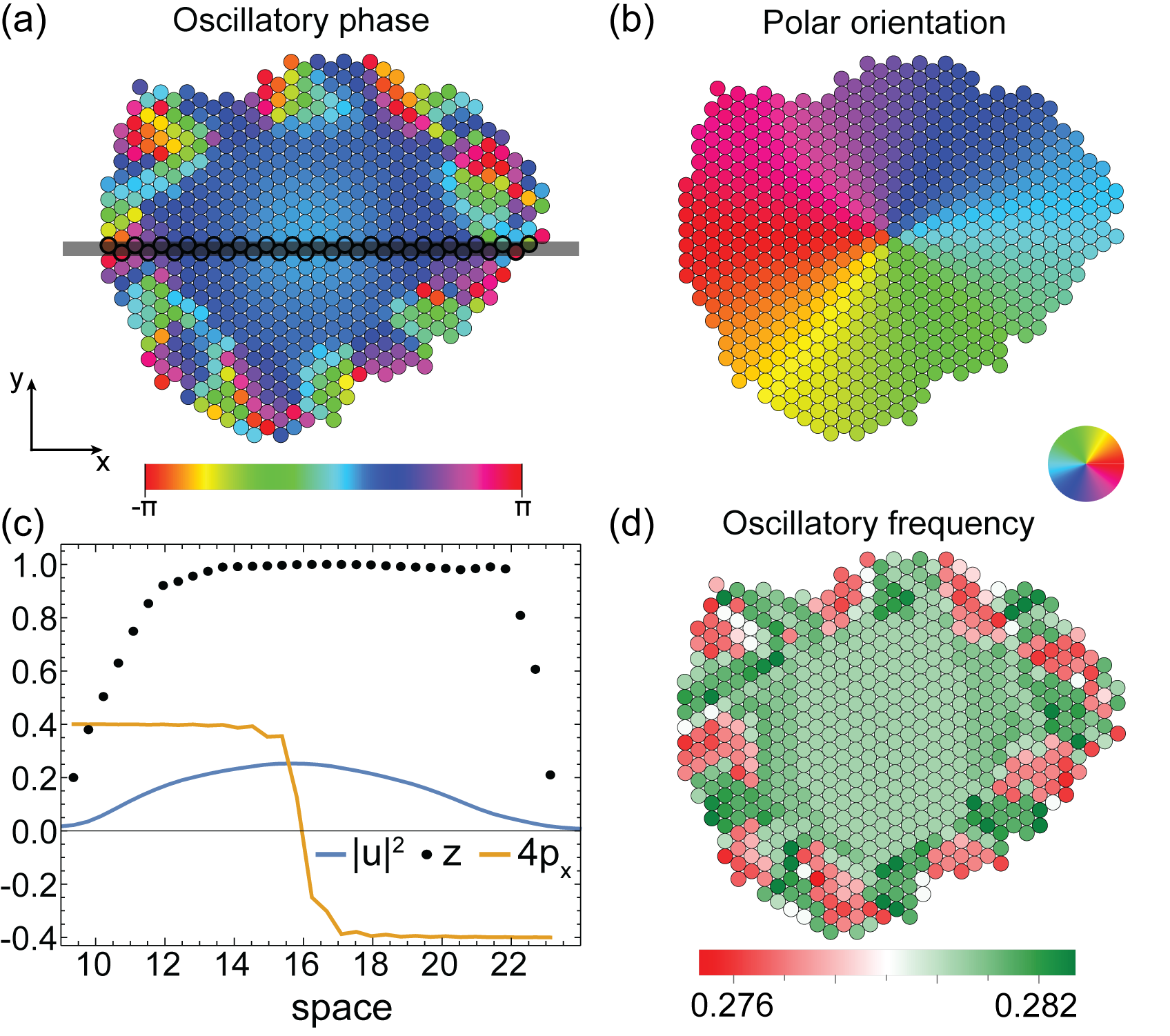}
    \caption{\textbf{Detailed analysis of volvox state.} (a) Agents' oscillatory phases, Eq.~\eqref{app_eq:phases}. (b) Polar orientation of agents. (c) Quantification of acoustic field amplitudes $\lvert u\rvert$, local Kuramoto order parameter $z$, Eq.~\eqref{app_eq:Kuramoto_local}, and polar orientation in $x$ direction $p_x$ for agents along the gray shaded line indicated in panel (a). (d) Oscillatory frequencies. Parameters as stated in Table \ref{tab:partPars} for Fig.~\ref{fig:pd}.}
    \label{app_fig:volvox}
\end{figure}

As the final distinct phenotypical states, we consider volvoxes. These exhibit a polar orientation toward a central defect, Fig.~\ref{app_fig:volvox}(b), reminiscent of blob solutions but lacking full synchronization. In the surrounding region, oscillators display uncorrelated phases or decoupled phase waves, Fig.~\ref{app_fig:volvox}(a), as coupling to the central oscillations via the acoustic field is too weak to achieve full synchronization.
The presence of these phase waves is further indicated by a reduced Kuramoto order parameter in the periphery, Fig.~\ref{app_fig:volvox}(c). Compared to the acoustic amplitudes in blob solutions, Fig.~\ref{app_fig:blob}(c), volvoxes exhibit greater spatial variation in amplitude between the aggregate’s core and the outer ring of uncorrelated agents. Regarding the oscillatory frequencies, the synchronized central fraction of agents exhibits the highest values, while those at the perimeter display a broader distribution of frequencies, Fig.~\ref{app_fig:volvox}.

\subsection{Data underlying quantitative phase diagram}\label{app:pd}
\begin{figure}[!ht]
    \centering
    \includegraphics[width=\linewidth]{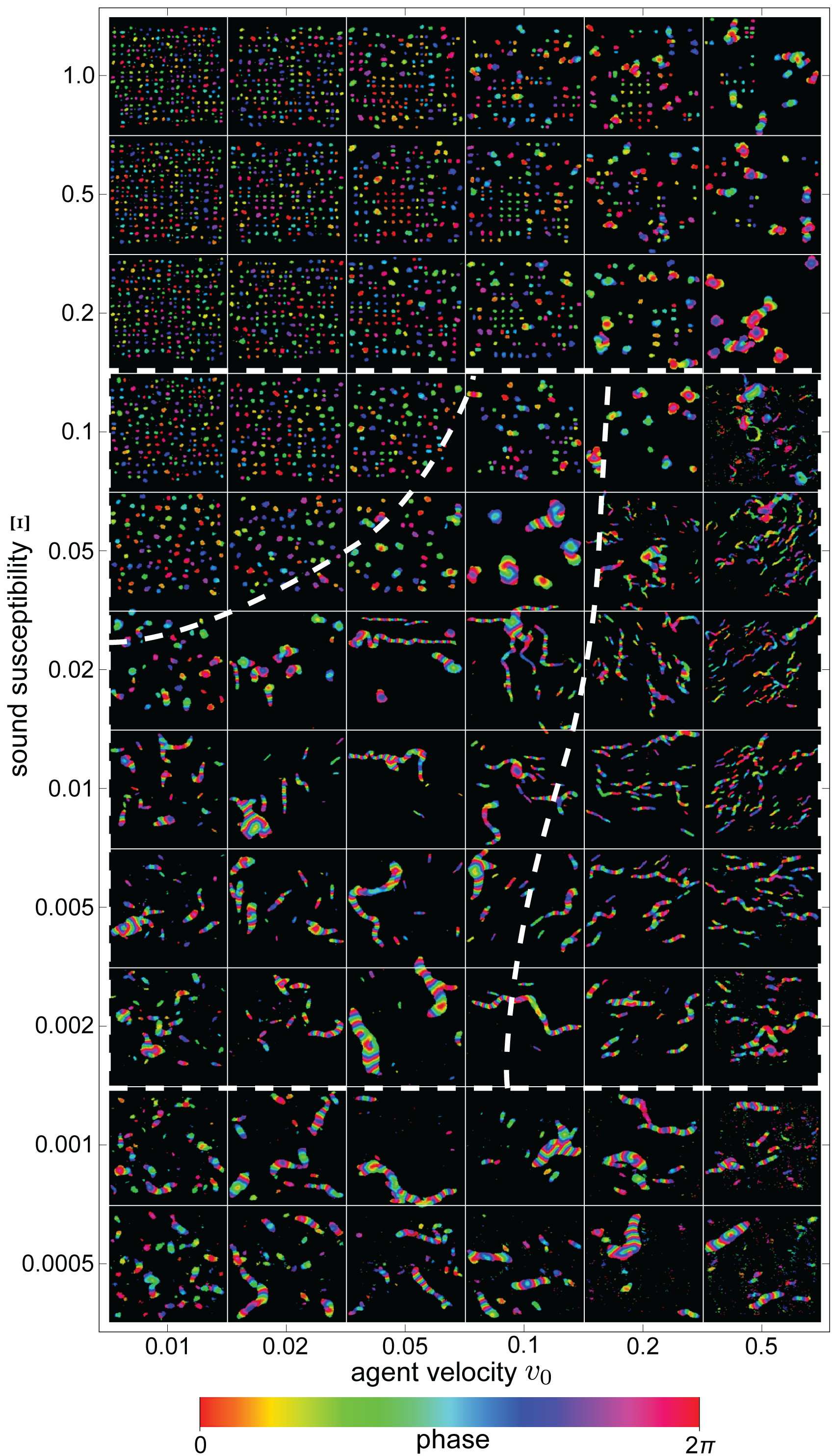}
    \caption{\textbf{Simulation data of phase diagram.} Final states of numerical simulations underlying the phase diagram, Fig.~\ref{fig:pd}. Agent color represents its oscillatory phase. The transitions between predominantly occurring phenotypical states (white dashed lines) correspond to the quantitative analysis of the numerical solutions, see section \ref{sec:pd_quant}. Each tile represents a numerical simulation involving $N=5000$ agents on a square domain with side length $L=100$. Numerical simulation parameters as stated in Table \ref{tab:partPars} for Fig.~\ref{fig:pd}.}
    \label{app_fig:quant_pd}
\end{figure}

In addition to the clustering measure $\Psi_\text{clust}$, Eq.~\eqref{eq:clustering}, and the average cluster polar order parameter $\Psi_\text{pol}$, Eq.~\eqref{eq:order_pol}, used to obtain the transition lines between predominant phenotypes, Fig.~\ref{fig:pd}, we provide the underlying simulation data in Fig.~\ref{app_fig:quant_pd}. We observe regions with the distinct phases; blobs are predominant in the top left corner for low agent velocities and larger sound susceptibilities. At intermediate parameter ranges, we observe the formation of larva structures, whereas at higher sound susceptibilities, the larvae exhibit phase defects in their head that can lead to ouroboroi formation. For high agent velocities, snakes are the predominant collective state in the system.

In addition, we observe various instances of multistability at the transition lines between predominant phenotypical collective states, Fig.~\ref{app_fig:quant_pd}. For example, at the transition between blob and larva states, numerical simulations reveal the coexistence of blobs and slowly moving larvae. Similarly, both larva and snake structures coexist close to their respective transition line. Moreover, within the regions characterized by the qualitatively distinct phenotypes, variations in collective behavior persist. Blobs tend to aggregate into larger clusters as the system approaches the transition to larva dominance, driven by increasing agent velocity $v_0$ or decreasing sound susceptibility $\Xi$. Larvae become increasingly symmetric with higher $\Xi$, while snakes stronger elongate for larger $v_0$. These observations indicate that transitions between different collective states do not occur sharply but rather over extended regions of parameter space, where minor behavioral variations persist even within a given phenotypical class.

\section{Parameter dependence of observed states}\label{app:dependence}
In the main text, we discuss the effects of key parameters in the presented model of acoustic active matter. As shown in the qualitative phase diagram (Fig.~\ref{fig:pd}) as well as in its quantitative analysis (Fig.~\ref{fig:quant_pd}), the interplay between attraction to the acoustic field, as mediated by the acoustic susceptibility $\Xi$ and the self-propulsion of agents, here controlled by their self-propulsion velocity $v_0$, gives rise to transitions between dominance of different phenotypical collective states.  Even though quite generally, the determinant factor for the predominantly occurring phenotypes is given by the interplay between self-propulsion effects and the response of collective solutions to the acoustic signaling, we present some additional quantification and insights into how other model parameters impact the observed behavior.

To better assess the differences in acoustic signaling and synchronization behavior upon parameter changes, we quantify the dependence of the agents' collective oscillatory states by means of the cluster-averaged Kuramoto order parameter \cite{KuramotoChemical},
\begin{align}\label{app_eq:Kuramoto}
    \Psi_{\text{phase}}=\frac{1}{N}\bigg\lvert\sum_\text{clusters}\sum_{j\in\text{cluster}}e^{i\phi_j}\bigg\rvert\,.
\end{align}
This order parameter quantifies the average synchronization of oscillatory phases $\phi_j$ within a cluster.
In the following we explore the behavioral changes of the emergent collective states in dependence on the polar alignment rate $\Gamma$, the acoustic coupling strength $\lambda$, the strength of rotational noise $\xi$, and the non-linear frequency coupling $b$, in numerical simulations of the agent-based model, Eqs.~\eqref{eq:agents}.

\subsection{Polar alignment rate $\Gamma$}
\begin{figure}
    \centering
    \includegraphics[width=\linewidth]{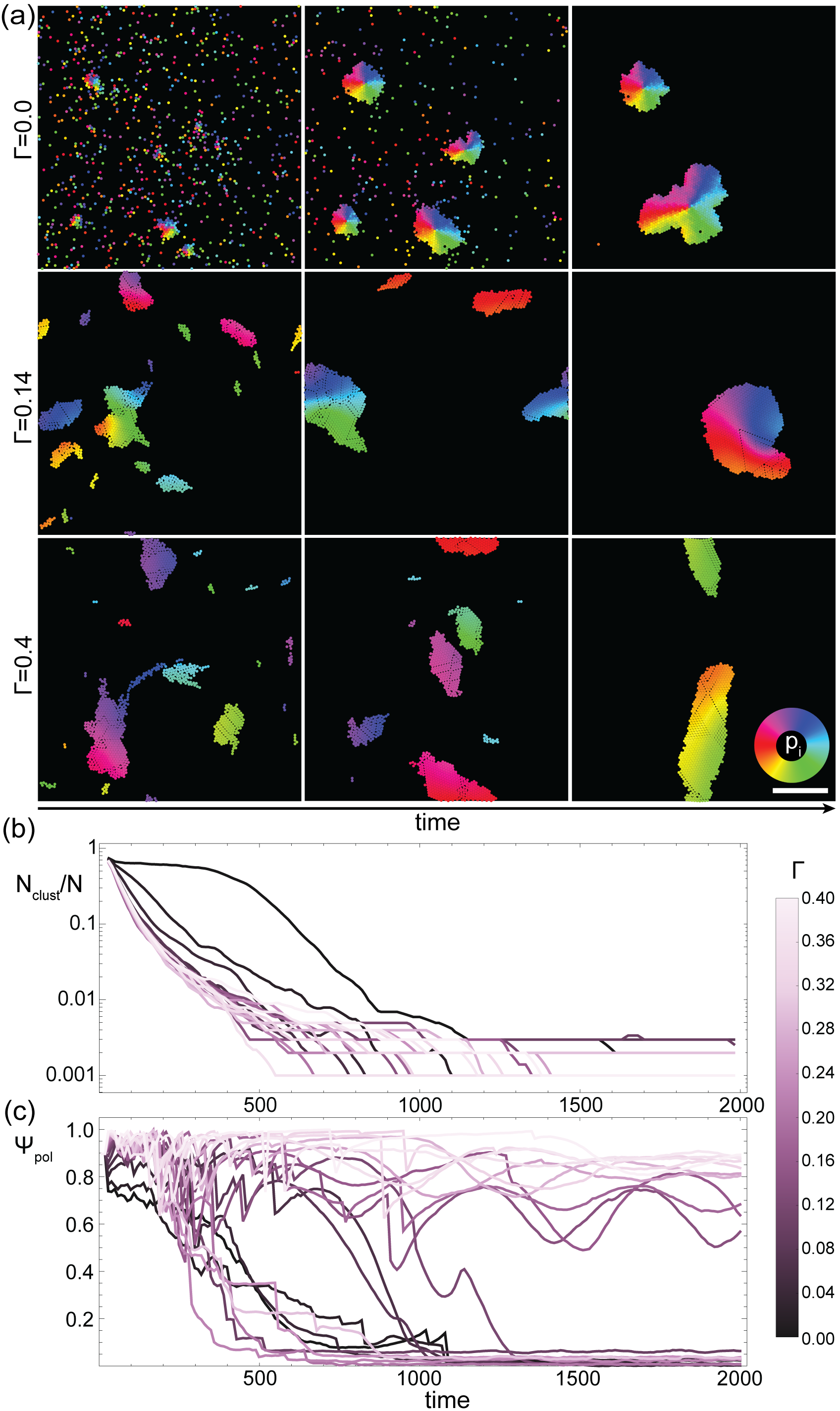}
    \caption{\textbf{Dependence of aggregation behavior on polar alignment between agents.} (a) Temporal evolution of the aggregation process for different values of the polar alignment parameter $\Gamma\in\left\{0.0,0.14,0.4\right\}$. Colors show the agents' polar orientation $\bm{p}_i$, and the white scale bar indicates a length of 10 units. (b,c) Temporal evolution of cluster number $N_\text{clust}$ (b) and average cluster polar order $\Psi_\text{pol}$ (c), for different values of polar alignment parameter $\Gamma$ (color code). Parameters as stated in Table \ref{tab:partPars} for Fig.~\ref{fig:pd} with $\Xi=0.05$, $\omega_u=0$, and $v_0=0.1$.}
    \label{app_fig:polar_align}
\end{figure}

Firstly, we address the role of the polar alignment rate $\Gamma$ for the emergent self-organization process. Even in the absence of polar alignment between the agents ($\Gamma=0$, Fig.~\ref{app_fig:polar_align}), aggregation still occurs due to acoustically mediated synchronization, which leads to the formation of regions with higher acoustic amplitudes. In this case, agents adjust their direction of motion solely based on the strongest acoustic signals which results in the emergence of localized, blob-like clusters.
By analyzing the temporal evolution of the number of clusters $N_\text{clust}$, we observe a slower aggregation process when polar alignment between the agents is absent (Fig.~\ref{app_fig:polar_align}(b)). In contrast, introducing polar alignment ($\Gamma > 0$) facilitates directed collective motion toward self-organized aggregation centers, accelerating the merging of clusters. In accordance with this observation, also the average polar order of aggregates, $\Psi_\text{pol}$, increases with the alignment strength (Fig.~\ref{app_fig:polar_align}(c)).
For higher polar alignment rates $\Gamma$, the system initially exhibits small, snake-like structures during the early stages of the dynamics. Over time, these structures evolve into larger aggregated snake structures that dominate the long-term behavior, Fig.~\ref{app_fig:polar_align}(a) ($\Gamma=0.4$). Thus, polar alignment plays a crucial role in the emergence of the elongated snake- and larva- structures. It enhances the initial aggregation process and shapes the system’s long-term dynamics toward final collective states with higher cluster polar order parameters $\Psi_\text{pol}$.

\subsection{Acoustic coupling strength $\lambda$}
\begin{figure}
    \centering
    \includegraphics[width=\linewidth]{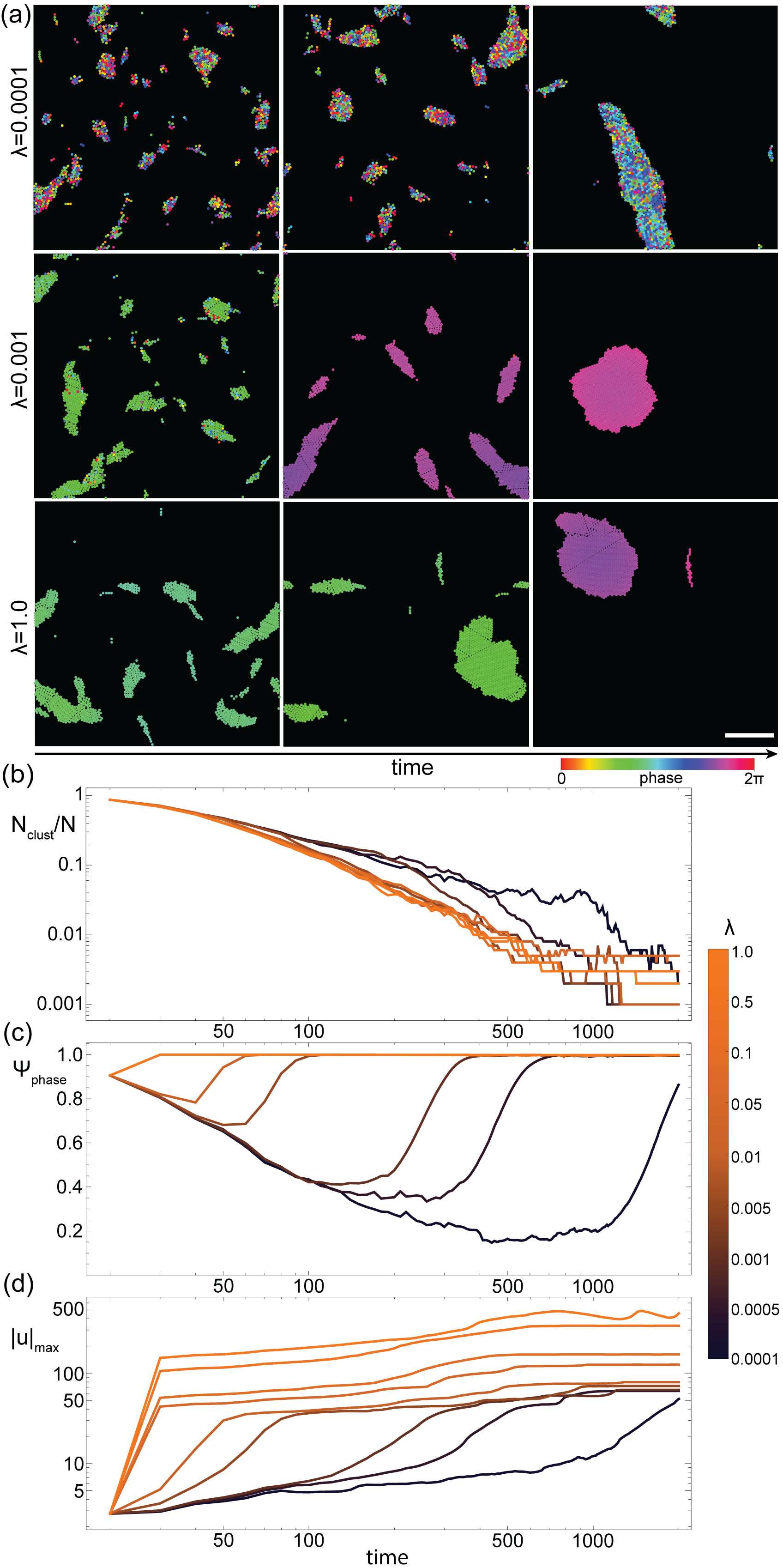}
    \caption{\textbf{Dependence of aggregation behavior on acoustic coupling between agents.} (a) Temporal evolution of agent-based simulations of Eqs.~\eqref{eq:agents} at different values of acoustic coupling strength $\lambda$. Temporal evolution of the normalized number of clusters $N_\text{clust}$ (b), cluster averaged Kuramoto order parameter $\Psi_\text{phase}$, Eq.~\eqref{app_eq:Kuramoto} (c), and maximum of acoustic field amplitudes $\lvert u\rvert_\text{max}$ (d) for different $\lambda$ (color code). Parameters as stated in Table \ref{tab:partPars} for Fig.~\ref{fig:pd} with $\Xi=0.05$, $\omega_u=0$, $v_0=0.1$, and $\Gamma=0.1$.}
    \label{app_fig:acoust_coupling}
\end{figure}

The acoustic coupling strength, $\lambda$, is another key parameter that influences the system’s behavior, as it determines the impact of the collective acoustic field on individual oscillations. When $\lambda=0$, agents are completely decoupled. However, increasing the coupling strength enhances synchronization through acoustic interactions. At weak coupling, $\lambda = 0.0001$, agents still aggregate into collective states, but even within clusters, the agents remain partially desynchronized over long timescales (Fig.~\ref{app_fig:acoust_coupling}). As a result, the number of clusters, $N_\text{clust}$, decreases more slowly compared to cases with stronger coupling, Fig.~\ref{app_fig:acoust_coupling}(b). Additionally, the cluster Kuramoto order parameter, $\Psi_\text{phase}$, and the maximum acoustic amplitudes, $\lvert u\rvert_\text{max}$, remain low for extended periods (Fig.~\ref{app_fig:acoust_coupling}(c,d)), leading to weaker attraction between agents.
When synchronization within a cluster is low, $\Psi_\text{phase}\lesssim 0.8$, as observed for $\lambda=0.0001$, the resulting weak acoustic fields are insufficient to sustain persistent localization of agents. Consequently, the system exhibits snake-like collective states. In contrast, for $\lambda>0.001$, agents achieve high synchronization, $\Psi_\text{phase}>0.95$, Fig.~\ref{app_fig:acoust_coupling}(b), and acoustic field amplitudes are significantly stronger, Fig.~\ref{app_fig:acoust_coupling}(c). This leads to the formation of localized blob states.
It is important to note that in this study we do not consider variations in the agents' internal frequencies. Such heterogeneity could further counteract synchronization and thereby select snake states over localized blobs for larger regions of the parameter space.
Overall, as $\lambda$ increases, stronger acoustic interactions promote faster synchronization of agents. This leads to higher acoustic signal amplitudes and, in turn, accelerates aggregation and reinforces the agents’ attraction toward regions of high acoustic activity.

\subsection{Rotational noise strength $\xi$}
\begin{figure}
  \includegraphics[width=\linewidth]{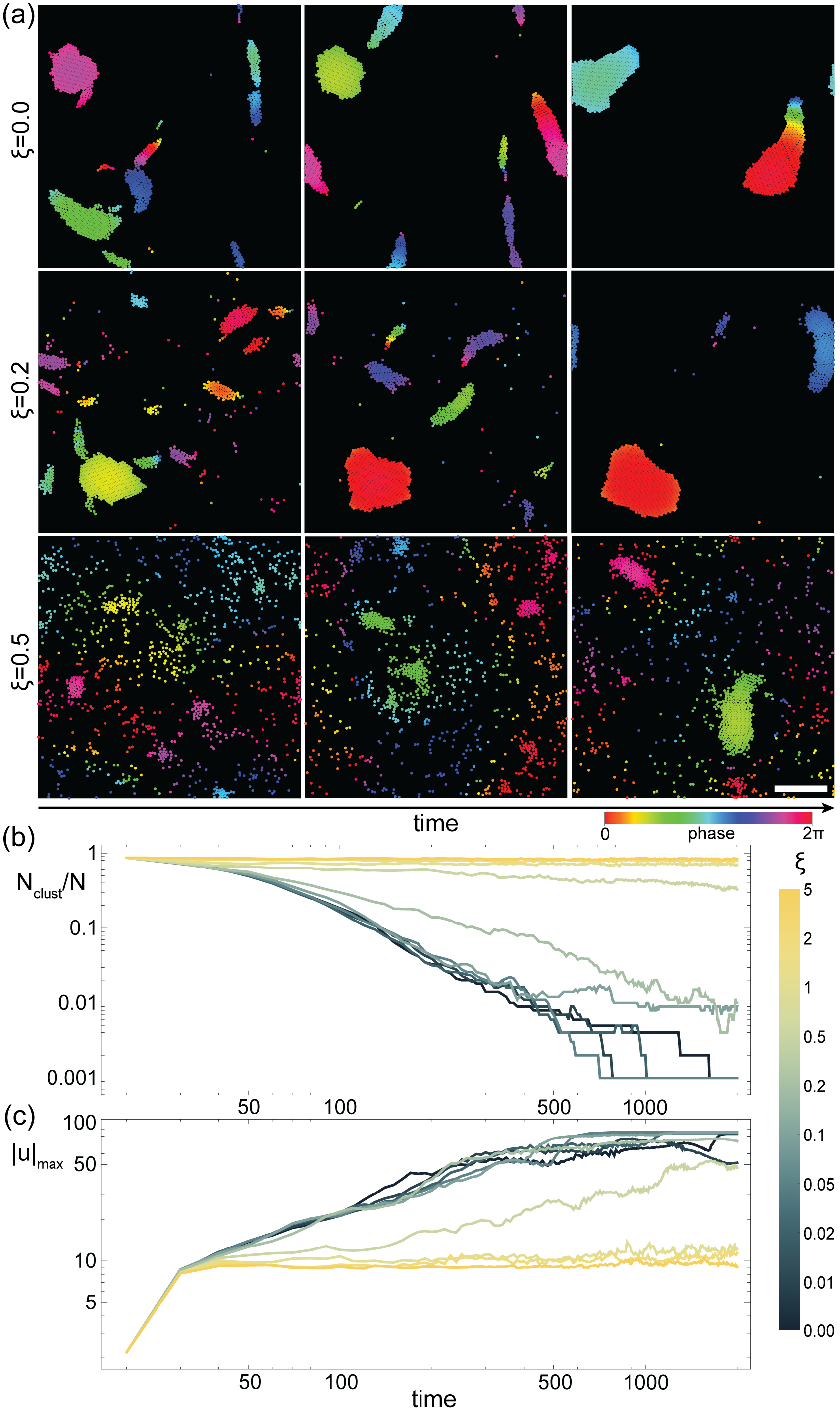}
    \caption{\textbf{Dependence of aggregation behavior on varying noise strength} $\xi$\textbf{.} (a) Temporal evolution of agents at different strengths $\xi$ of rotational noise. Dynamics of number of clusters $N_\text{clust}$ (b), and maximum acoustic field amplitudes $\lvert u\rvert_\text{max}$ (c) for different $\xi$ (color code). Parameters as stated in Table \ref{tab:partPars} for Fig.~\ref{fig:pd} with $\Xi=0.05$, $v_0=0.1$, $\omega_u=0$, and $\Gamma=0.1$.}
    \label{app_fig:noise_strength}
\end{figure}

Another factor that influences the persistence of the agents' directed motion is the rotational noise strength $\xi$. In the absence of rotational noise, $\xi=0$, agents align perfectly in a polar fashion with their neighbors and toward acoustic aggregation regions, Fig.~\ref{app_fig:noise_strength}. As a result, low noise values correspond to fast aggregation processes, Fig.~\ref{app_fig:noise_strength}(b). Due to the efficient and persistent aggregation, agents synchronize more effectively, which leads to large acoustic field amplitudes, Fig.~\ref{app_fig:noise_strength}(c).
As rotational noise increases, e.g., $\xi=0.2$, a growing number of agents experience stronger stochastic reorientations, preventing them from permanently aggregating. The attraction toward regions with high sound amplitudes within the aggregates becomes insufficient to counteract these random reorientations, causing an increasing fraction of agents to remain in the non-aggregated gas phase.
For strong noise levels, $\xi\gtrsim 1$, spontaneous reorientations completely dominate over acoustic attraction. Under these conditions, and for the chosen agent density, aggregation is entirely suppressed, preventing the formation of stable clusters, $N_\text{clust}\sim N$, Fig. \ref{app_fig:noise_strength}(b).

\subsection{Nonlinear frequency coupling $b$}
\begin{figure}
  \includegraphics[width=\linewidth]{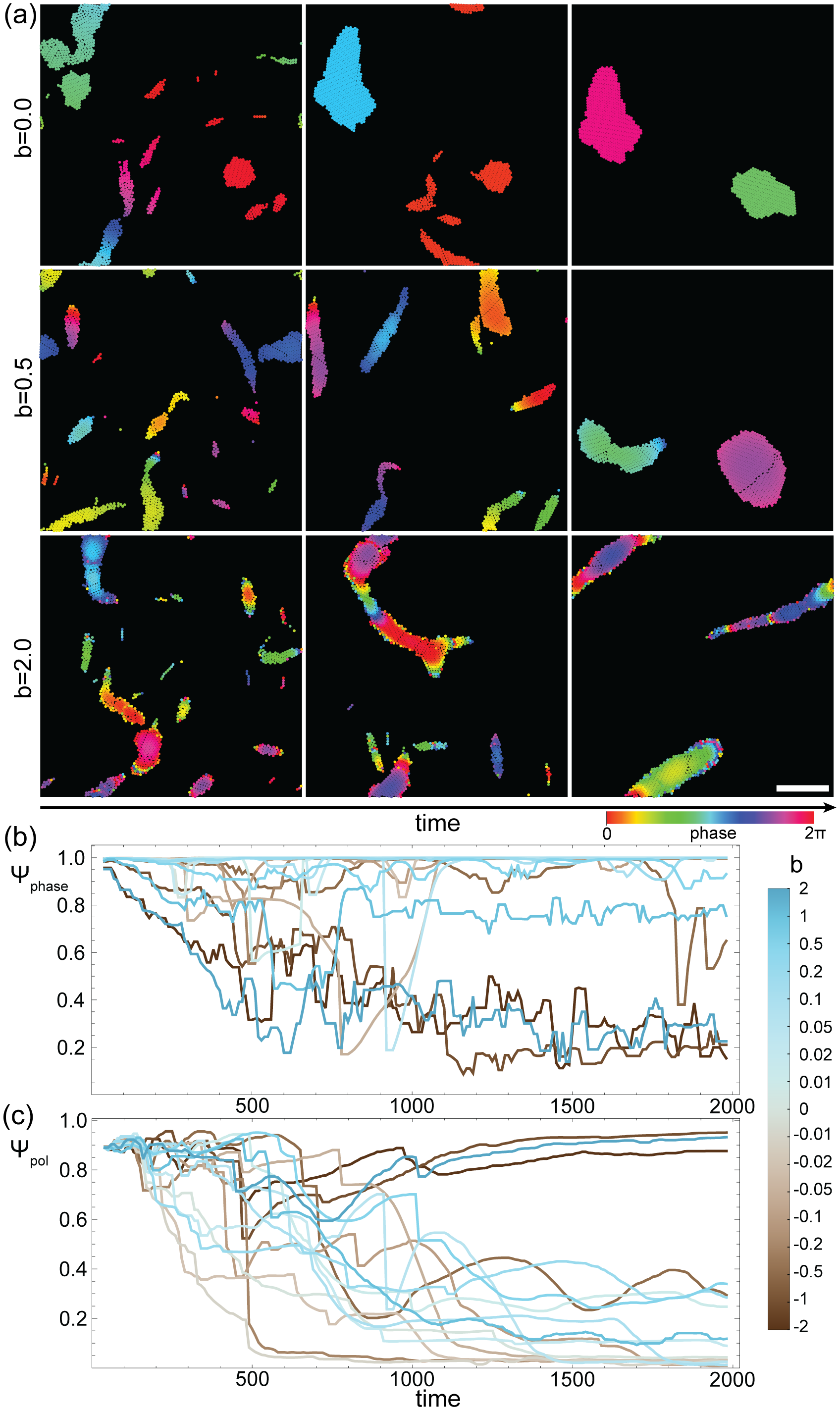}
    \caption{\textbf{Dependence of aggregation behavior on nonlinear frequency coupling.} (a) Temporal evolution of agents at different coupling values $b\in\left\{0,0.5,2\right\}$. Dynamics of the cluster Kuramoto order parameter $\Psi_\text{phase}$, Eq.~\eqref{app_eq:Kuramoto} (b), and cluster polar order $\Psi_\text{pol}$ (c) for different $b$ (color code). Parameters as stated in Table \ref{tab:partPars} for Fig.~\ref{fig:pd} with $\Xi=0.05$, $\omega_u=0$, and $v_0=0.1$.}
    \label{app_fig:nl_coup}
\end{figure}

In the dynamic equations for the oscillatory units, frequencies are nonlinearly coupled to the oscillators' amplitudes by the parameter $b$. Here, we discuss the impact of this coupling on the observed collective behavior, Fig.~\ref{app_fig:nl_coup}.
In the absence of a nonlinear frequency coupling, $b=0$, oscillatory frequencies completely decouple from the amplitudes of the individual oscillators and, thus, also from the overall local acoustic field amplitudes. For this case, the system is no longer able to establish phase leaders, and the formation of asymmetric polar defect positions, such as in larvae states, is harder to achieve. Clusters fully synchronize their phases, $\Psi_\text{phase}\approx 1$ (Fig.~\ref{app_fig:nl_coup}(b)), and we observe the formation of localized blobs. In turn, for a non-vanishing frequency coupling, $b>0$, agents do not fully synchronize. Phase-leaders and phase waves emerge within the aggregates and polar motion in the form of larvae, ($0.2\lesssim\Psi_\text{pol}\lesssim 0.8$) and snakes ($\Psi_\text{pol}>0.8$) become predominant ($b=2.0$ in Fig.~\ref{app_fig:nl_coup}).

\section{Details on numerical simulations}\label{app:num}
\subsection{Agent-based simulations}
In this section, we give the details of the applied numeric integration scheme for the dynamics of the $N$ acoustically interacting active agents in a two-dimensional domain $(x,y)\in\left[0,L\right]\times\left[0,L\right]$ with periodic boundary conditions.
In the agent-based simulations, we directly solve the equations \eqref{eq:agents} for the agents' positions $\bm{r}_l$, orientation angles $\varphi_l$, and oscillatory states $a_l$ applying a forward Euler-Maruyama scheme with fixed time step $dt$. Agents interact through a hard-sphere interaction rule $f_{lj}$, causing overlapping agents to shift in the direction of their distance vector equally until a distance of $2\,r_{\text{p}}$ is restored. Further, agents exhibit polar alignment with neighboring agents within an interaction radius of $r_c = 4\,r_{\text{p}}$. We assume a stochastic reorientation of individual agents by adding a zero-mean Gaussian white noise with amplitude $\xi_l$ to the dynamics of the orientation angle. For the direct agent-agent interactions, neighborhoods are tracked using regular linked cell lists with periodic wrapping.
To implement the acoustic interactions between swarmers, we compute the instantaneous quasi-stationary acoustic field in Fourier space as detailed in section \ref{sec:model_agents}. Each agent contributes as a source according to its real unbinned position with its current oscillator state $a_l$, extended as a Gaussian contribution with width $\sigma=2\,r_{\text{p}}$. Namely, the source contribution of agents to the acoustic field is modeled as
\begin{align}
    w(\bm{r})=\alpha\exp\left\{-\bm{r}^2/2\sigma^2\right\}\,,\quad\text{with }\sigma=r_\text{p}\,,
\end{align}
and constant amplitude scale $\alpha=50$.
Depending on the size of the studied system, we resolve the acoustic field with $128$ up to $512$ Fourier modes per spatial direction. Derivatives of the acoustic field amplitudes with which the swarmers align their direction of self-propagation are computed in Fourier space. Large aggregates of agents act as spatially extended sources, collectively emitting acoustic waves with a spatial extension limited only by the system size. With the chosen implementation of acoustic waves through a quasi-stationary solution, Eq.~(\ref{eq_app:a_solution}), and neglecting absorption, large-scale patterns are overemphasized, and divergencies of the Fourier kernel $\sim (k_x^2+k_y^2)^{-1/2}$ require regularization. To address this uncontrolled cutoff, we introduce a parameter,  $k_{\text{crit}}$, which imposes an upper bound on the wavelength, $\ell = 2\pi/k_{\text{crit}}$, of acoustic wave patterns. Throughout this study, we set $k_{\text{crit}} = 1$. Future studies may go beyond this effective approach by incorporating acoustic damping and accounting for acoustic wave dynamics to include effects such as Doppler shift.
If not stated otherwise, the swarmers are initially uniformly distributed in the two-dimensional domain, avoiding overlaps between swarmers when drawing random positions. Their orientation angles are picked from a uniform distribution, $\varphi_l\in\left[0,2\pi\right]$ and oscillatory states are set to the stable amplitude $\lvert a\rvert=1$ with uniformly distributed phases $\left[0,2\pi\right]$.

\subsection{Continuum field equations}
In this appendix section, we give details on the numerical solution of the continuum field model for acoustic active matter, Eqs.\@ \eqref{eq:field}. We consider the equations on a 2D square domain, $\bm{r}=(x,y)^T\in\left[0,L\right]\times\left[0,L\right]$. 
The pressure-like contribution to the polar field dynamics
\begin{align}
    P'(\rho)&=\exp\left(-32\rho\right)+\exp\left(16(\rho-2)\right)\,,
\end{align}
implements the minimal and maximal agent densities at the values zero and two, respectively.
We numerically solve the equations for the swarmer density $\rho(\bm{r},t)$, oscillator state $a(\bm{r},t)$, and polar orientation $\bm{p}(\bm{r},t)$ in Fourier space by applying pseudo-spectral methods with typical resolutions of 256 to 8192 modes per dimension, depending on system size. The critical computations, and particularly, the fast Fourier transform are parallelized running on the graphical processing unit (GPU) using NVIDIA CUDA computing platform. We implement differential operators using their respective Fourier kernels, and nonlinearities in the dynamics are computed in real space. The temporal integration is implemented using exponential time differencing (ETD2), realizing a semi-implicit solving of the linear contributions to the dynamics. As for the agent-based simulations, we employ the stationary solutions for the acoustic field $u$, Eq.\@ \eqref{eq:sound_sol}, with continuous sources $g(\bm{r},t)=\rho(\bm{r},t)\,a(\bm{r},t)$. The system is typically initialized with a homogeneous density of $0.6$, well below the isotropic to polar order transition at $\rho=1$, perturbed with a zero-mean Gaussian white noise of small amplitude. We initialize the polar order field as well as swarmer's complex oscillatory state with zero-mean Gaussian white noise around vanishing values.

\section{Model parameters}
\label{app:model_parameters}
The system parameters used for the numerical simulations of the agent-based model, Eq.~(\ref{eq:agents}) are given in Table \ref{tab:partPars}.
\begin{table*}[!tb]
\centering
\begin{tblr}{colspec={|c|c|c|c|c|c|c|c|c|c|}, colsep=4pt}
			\hline
			Parameter & Description 
			&{Fig.\@ \ref{fig:aggregation} } &{Fig.\@ \ref{fig:pd} } &{Fig.\@ \ref{fig:sensing}(a) } &{Fig.\@ \ref{fig:sensing}(b) } &{Fig.\@ \ref{fig:function}(a) } &{Fig.\@ \ref{fig:function}(b) } &{Fig.\@ \ref{fig:function}(c) }&{Fig.\@ \ref{fig:function}(d)}  \\
			\hline[1pt]
			$v_0$  &  agent velocity    & 0.1&   various &  0.1 &  0.1  &  0.2 &  0.1 & 0.2 &0.1   \\ \hline
			$D_{\rm R}$ & angular noise & 0.02 &0.02 & 0.02 &  0.01 &   0.01 &  0.02 & 0.01 &0.01    \\ \hline
			$r_{\text{p}}$ & particle radius & \SetCell[c=8]{c}{$0.25$} \\
			\hline
            $r_c$  &  interaction radius  &
            \SetCell[c=8]{c}{$4\,r_{\text{p}}$}  \\ \hline 
			$\Gamma$ & polar alignment factor & 0.2 &  0.1 & 0.1&  0.2 &  0.2 &  0.1 & 0.2 & 0.2 \\
			\hline
   $\Xi$ & sound susceptibility &0.05 &various & 0.05&  0.1 &    0.01 &  0.05 &  0.2 & 0.05 \\
			\hline
   $\omega$ & free oscillator frequency&0.5 & 0.5 & 0.1&  0.5 &   0.5 &  0.1 &  0.5 &0.5 \\
			\hline
            $\omega_u$  &  field frequency  &
            0&\SetCell[c=7]{c}{$\omega$}  \\ \hline 
   $b$ & nonlinear frequency coupling &0.5&0.5 & 0.05&  0.05 &    0.5 &  0.05 &  0.05 & 0.05 \\
			\hline
   $\lambda$ & acoustic coupling &0.005& 0.1 & 0.1&  0.02 &   0.00001 &  0.1 &  0.01 & 0.1 \\
			\hline
   $c$ & sound speed &20&20 & 5&  50 &    50 &  50 &  50 &50 \\
			\hline
\end{tblr}
\caption{Parameters used for the agent-based model, Eqs.\@ \eqref{eq:agents}. }
		\label{tab:partPars}
\end{table*} 
For the numerical simulations of the continuous field equations, Eqs.\@ \eqref{eq:field}, we used the parameter values detailed in Table \ref{tab:fieldPars}.
\begin{table*}[!tb]
\centering
\begin{tblr}{colspec={|c|c|c|c|c|c|}, colsep=4pt}
			\hline
			Parameter & Description 
			&{Fig.\@ \ref{fig:field} } & {Fig.\@ \ref{fig:field_sound}} &{Fig.\@ \ref{fig:field_coarsen}}& {Fig.\@ \ref{fig:defect_coarsening} }  \\
			\hline[1pt]
			$v_0$  &  agent velocity     &  various& 0.02 (a), 0.1 (b)& 0.04  & 0  \\ \hline
			$\mu$ & density diffusion & 0.05 (a), 0.01 (b) & 0.05 (a), 0.02 (b)& 0.02&0  \\ \hline
			$\sigma$ & polar order transition factor& 1 & 1&1& n.a. \\ \hline
            $\delta$  &  polar order magnitude parameter  &   $1$& 1&1  & n.a. \\ \hline 
			$\kappa$ & polar elasticity & 0.05 & 0.05&0.05& n.a. \\ \hline
              $\chi$ & polar self-advection & 0.05 & 0.05&0.05& n.a.  \\ \hline
            $\Xi$ & sound susceptibility & various & 5000 (a), 100 (b)&100& n.a. \\ \hline
   $\omega$ & oscillation frequency & 0.5 & 0.5&0.5& 0.5 \\ \hline
   $b$ & nonlinear frequency coupling & 0.05 & 0.05&0.05& 0.05\\ \hline
   $\lambda$ & acoustic coupling & 0.001 (a), 1 (b) & 1000 (a), 0.001 (b)&1& various\\	\hline
   $c$ & sound speed & 50  & 50 (a), 20 (b) &50& 50\\	\hline
\end{tblr}
\caption{Parameters used for numerical simulations of the continuous field equations, Eqs.\@ \eqref{eq:field}. }
		\label{tab:fieldPars}
\end{table*} 

\FloatBarrier

\clearpage\newpage

\end{document}